\documentclass[review]{elsarticle}

\usepackage{amsmath}
\usepackage{amsfonts}
\usepackage{amssymb}
\usepackage{graphicx}
\DeclareGraphicsExtensions{.eps}
\usepackage{subcaption}
\usepackage{multirow}
\usepackage{breqn}
\usepackage{placeins}
\usepackage{makecell}
\usepackage{hyperref}
\usepackage{algorithm2e}
\usepackage[usenames, dvipsnames]{color}
\usepackage{url}
\journal{Journal of \LaTeX\ Templates}

\begin{document}

\begin{frontmatter}

\title{Across-scale Process Similarity based Interpolation \\ for Image Super-Resolution
}


\author{Sobhan Kanti Dhara\corref{mycorrespondingauthor}\cortext[mycorrespondingauthor]{Corresponding author}}
\ead{dhara.sk@gmail.com}
\author{Debashis Sen}
\ead{dsen.eece.iitkgp@gmail.com}
\address{Department of Electronics and Electrical Communication Engineering \\ Indian Institute of Technology, Kharagpur, India}
\begin{abstract}
 A pivotal step in  image super-resolution techniques is interpolation, which aims at generating high resolution images without introducing artifacts such as blurring and ringing. In this paper, we propose a technique that performs interpolation through an infusion of high frequency signal components computed by exploiting `process similarity'.  By `process similarity', we refer to  the resemblance   between a decomposition of the image at a resolution to the decomposition of the image at another resolution. In our approach, the decompositions generating image details and approximations  are obtained through the   discrete wavelet  (DWT) and stationary wavelet  (SWT) transforms. The complementary nature of  DWT and  SWT is leveraged to get the structural relation between the input image and its low resolution approximation. The structural relation is represented by optimal model parameters obtained through particle swarm optimization (PSO).   Owing to process similarity, these  parameters are  used to generate the high resolution output image from the input image.  The proposed approach is compared with six   existing techniques qualitatively and in terms of PSNR, SSIM, and FSIM  measures, along with computation time (CPU time).   It is found that our approach is the fastest in terms
of CPU time and produces comparable results.
\end{abstract}


\end{frontmatter}
\section{Introduction}
\label{Introduction}
Due to constraints in device and sensor technology, an image may not be captured in the required high resolution \cite{Park2003}. Moreover, for  high-speed transmission, the captured image is sometimes down-sampled to lower the bandwidth required \cite{wu2009low}. The transmitted low resolution image is then required to be processed at the receiver to generate the image in  higher resolution. Hence, approaches are needed to construct an image of the desired (higher) resolution from an image at lower resolution. Image super-resolution systems are designed for this purpose \cite{Park2003}. Super-resolution  has a critical role in applications where it is necessary to perform in-depth analysis of local regions for detection and recognition purposes \cite{yue2016image}.

Most super-resolution strategies involve interpolation as a pivotal step to increase the resolution of an image. But interpolation is prone to artifacts such as blurring and ringing effects. Blurring artifact is due to  attenuation at high frequencies which correspond to edges. The ringing artifact is due to the oscillation of signal at sharp edges because of the truncation of high frequency components \cite{Gonzalez2002}.   

Here, we discuss different interpolation techniques  starting from the  standard and well-known ones    to recently proposed techniques in  different categories  more or less in  a chronological order.  The standard and well-known interpolation techniques are bilinear, bicubic  and Lanczos, which introduce artifacts, especially at the edges \cite{Keys1981, Wolberg1990}. Due to computational simplicity and  fast performance \cite{siu2012review}, these techniques  are used by the standard photo editing software like Photoshop\textsuperscript \textregistered \footnote{https://helpx.adobe.com/in/photoshop/using/resizing-image.html}. Moreover, in many  recently developed super resolution techniques, the input image  is     initially  interpolated \cite{kim2016accurate,kim2016deeply} or the chrominance channels are interpolated  \cite{Dong2013, yang2017deep}   using one of the basic  techniques.  To handle the artifacts in the interpolation, different edge directed interpolations are proposed in the literature \cite{Jensen1995, Li2001}. The technique proposed in \cite{Jensen1995} interpolates based on an edge fitting operator to maintain  visual integrity. The technique proposed in \cite{Allebach1996}  interpolates based on the direction of  edges, and consists of data correction and edge directed rendering operations. The new edge directed interpolation (NEDI)  proposed in \cite{Li2001} preserves low resolution covariance values at the higher resolutions based on resolution invariant property \cite{Li2001} of edge orientation.  Such a property   ensures interpolation along the edge direction avoiding blurring. NEDI is one of the widely referred techniques  in the domain of interpolation. 

 Discrete wavelet transform (DWT) \cite{mallat1999wavelet} decomposes an input image into four frequency subbands having low frequency component, and horizontal, vertical and diagonal edge components, separately. So, DWT based interpolation techniques in \cite{Carey1999a,nguyen2000wavelet,Demirel2011,Tian2011} can handle the edges efficiently.  Regularity preserving image interpolation proposed in \cite{Carey1999a} estimates the regularity of edges by measuring the decay of wavelet transform coefficients across scales. The algorithm also preserves the underlying regularity by extrapolating new subbands to be used in the generation of the high resolution image.  The DWT-based technique mentioned in \cite{Demirel2011} decomposes an input image at finer to coarser levels and the subbands generated  at each  level are interpolated for estimation. The estimated high
frequency subbands using DWT are then modified by using high frequency subbands obtained through the stationary wavelet transform (SWT) \cite{nason1995stationary}. Then, inverse DWT (IDWT) is applied to get a sharper high resolution image. Prediction of the wavelet coefficients and parameter optimization by ant colony optimization is proposed in \cite{Tian2011} for a DWT-based interpolation. In \cite{Tian2011}, a three-component exponential mixture (TCEM) model is used to analyze magnitude and sign information of wavelet coefficients. Ant colony optimization (ACO) \cite{dorigo2010ant} technique is used for optimal parameter estimation of the TCEM model.

Sparse representation model (SRM) for image super-resolution is being widely explored \cite{Zhang2008,Mallat2010, Yang2010, Dong2011,Dong2013}. The SRM based interpolation approach of \cite{Mallat2010}, sparse mixing estimator (SME),  interpolates by estimating a sparse image and  by mixing directional interpolators over oriented blocks in a wavelet frame.  SME computes mixing coefficients by minimizing $l^1$ norm which is weighted by signal regularity in each block. On the other hand, the sparse basis set-based technique of  \cite{Zhang2008,Dong2011} learns a map from a relevant training set. Thus the missing information is retrieved  or generated from training data. However,  in \cite{Ebrahimi2007}, the authors suggest  extraction of information from the input image  because it provides limited but  more relevant information than that from a universal database \cite{Freedman2011}. A self-exemplar SRM (without external training) is proposed in the  NARM-SRM-NL approach of \cite{Dong2013}, where nonlocal autoregressive modeling is used. The earlier autoregressive model   has been used \cite{Zhang2008} to exploit the local similarity  for  interpolation,  whereas, NARM-SRM-NL exploits both the local and nonlocal similarity in an adaptive manner and interpolates a pixel as a linear combination of its nonlocal neighbors.  Although the algorithm promises improved performance, it suffers from a huge computational burden due to patch clustering and PCA sub-dictionary learning.

Self-similarity based super-resolution (SR) algorithms generate higher resolution images by exploiting statistical prior. These algorithms  utilise the internal statistics of LR-HR patch pair \cite{Freedman2011, Singh2014,Huang2015}. The technique proposed in \cite{Freedman2011} searches extremely localized regions for the example patches. However, realistic reproduction of fine-detailed cluttered regions is not achieved,  and those regions appear somewhat faceted. Instead of searching for example patches from the input image, the technique proposed in \cite{Singh2014} searches  for patches in different subbands of the input image. Recently, in \cite{Huang2015},  the authors proposed a geometric patch transformation model which includes affine  and perspective transformations during interpolation by patch matching. However, patch matching and dictionary learning based algorithms are always computationally expensive.

Recently, few end to end deep neural networks have been proposed for super-resolution. SRCNN \cite{dong2014learning,dong2016image} uses the convolutional neural network (CNN)  \cite{krizhevsky2012imagenet} for image super-resolution. The technique is later extended to DRCN \cite{kim2016deeply} by increasing  the number of layers of the network and incorporating recursive learning.  The techniques proposed in \cite{kim2016accurate, yang2017deep}  use residual learning  in  deep  and very deep networks. EDSR \cite{lim2017enhanced} is an enhanced version of deep residual networks in single-scale architecture. The performance of EDSR is shown to be superior  compared to  other learning based super-resolution approaches. The super-resolution technique SRGAN \cite{ledig2017photo} uses the generative adversarial network (GAN) \cite{goodfellow2014generative} to generate perceptually high quality photo-realistic natural output images. As stated in \cite{Singh2014,cruz2018single}, such methods are well suited to perform on images of types similar to that in its training set.  

  Restoration of  edges   (which is  high frequency components) is crucial for the improvement of image visual quality \cite{shao2008edge, vishnukumar2014edge}.  Recently, authors in \cite{Acharya2017a}  proposed the so-called composite predictive technique to efficiently restore the high frequency and very high frequency components of the image. The technique extracts these components from the input low resolution image by exploiting the local statistics of an image. Later, in \cite{Acharya2017}, the authors  proposed a technique where fuzzy local adaptive Laplacian post-processing is used to enhance high frequency details in up-sampled images for interpolation. 
  
 The above discussion shows that image structure information at the desired high resolution image is related to that in the low resolution image. Such a relation is considered in many recent  interpolation approaches as structural similarity  \cite{deng2014structural}. A higher resolution image can be generated by modeling the relation between its structure information and that in its low resolution image (given input).  Such a modeling  may not assume structural similarity.  The relationship in
 structure information must be exploited, as it is the only relevant information 
about the high resolution image available  from the input image
 for constructing the high frequency details.

\begin{figure*}[!h]
\centering
\includegraphics[width=5.6in]{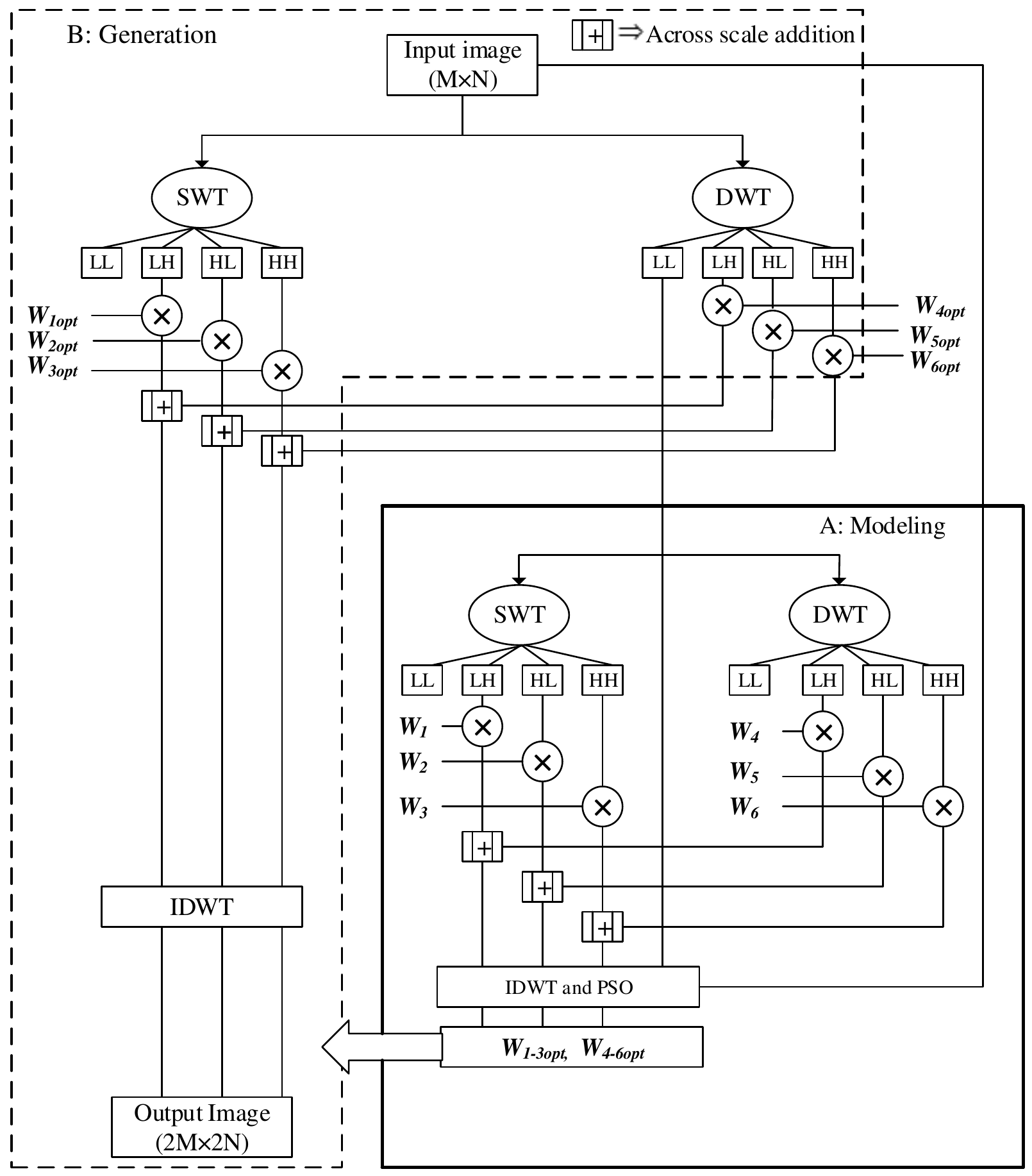}
\caption{Block diagram of the proposed interpolation approach for image super-resolution}
\label{fig_algo}
\end{figure*}

 In this paper, we propose a technique that performs interpolation by infusing   high frequency signal components computed leveraging `process similarity'.  For this, we consider the decomposition of an image at a higher resolution into a lower resolution image and high frequency details. Process similarity refers to the similarity between such a decomposition process of the image at  one resolution and that at another resolution. This process similarity allows us to model the relation between structure information of the input low resolution image and that of the desired high resolution image. The modeling can be based on the available relation between structure information of the input image and its low resolution approximation obtained by the decomposition (See Figure \ref{fig_algo}).  To accomplish this, discrete wavelet transform (DWT) and stationary wavelet transform (SWT) are used. The high frequency components and lower resolution approximations obtained through the said transforms are used to get the structural relation between the input image and its low resolution approximation. The structural relation is represented by model parameters obtained through particle swarm optimization (PSO).  The parameters  are then used to fuse the high frequency components that are subsequently employed to estimate the input image from its low resolution approximation.  The PSO works to minimize  the error between the actual input image and its estimate. Once the model parameters are obtained, they are then used to generate the desired high resolution image from the input image, just like the estimation process of input  image from its low resolution approximation. Note that, the structural relation modeling in our approach should not be considered as a modeling under the well-known concept of self-similarity \cite{yang2010exploiting} in super resolution. Presence of a structural relation between an image at two different resolutions does not necessarily mean that their structures are similar, that is, structural relation does not imply structural self-similarity.

 As mentioned earlier, the concept of process similarity allows modeling of structural relation based on the input image and its low resolution approximation. Therefore, the structural information used for interpolation is obtained from the input image itself, which is more likely to carry relevant information required to generate the image at higher resolution compared to a database of images  similar to the input image. Moreover, the high frequency components generated for the infusion are directional sensitive (vertical, horizontal and diagonal).  To incorporate  location consistency and value exactness (See  Section \ref{motimation}), our approach considers both DWT and SWT in the modeling.   These are the reasons why our 
technique  when compared to the existing, produces results having  lower artifacts, and  improves both the subjective and objective resemblance between the output high resolution  and the input low resolution images.  Finally, the  structural relation is represented by model parameters obtained through computationally inexpensive particle swarm optimization.  With respect to the algorithms compared, our approach is less expensive  in terms of memory and speed   requiring only   multi-scale decompositions up to two levels, multi-scale reconstruction  and   optimization using PSO. We consider comparisons of our approach based on visual inspection  as well as PSNR, SSIM, \cite{Wang2004} and FSIM \cite{Zhang2011}  measures with the widely accepted  edge directed NEDI \cite{Li2001}, nonlocal autoregressive modeling based NARM \cite{Dong2013}, recent  self-exemplar based SR-TSE \cite{Huang2015}, the latest fuzzy predictive composite scheme (FPCS-LAL) \cite{Acharya2017}, and deep learning based EDSR \cite{lim2017enhanced} and SRGAN \cite{ledig2017photo}. A preliminary version of this work has been published  in \cite{dhara2017}, where  a few other existing approaches were considered for result analysis. They, in general, perform inferior to the existing approaches considered here for comparison.  As discrete and stationary wavelet transforms, and particle swarm optimization are the core methods employed in our approach, a brief summary of them is provided in \ref{app1} and \ref{app2}.

 The overall contributions of the paper are as follows :

\begin{itemize}
\item We propose across-scale process similarity based on discrete wavelet transform (DWT) and stationary wavelet transform (SWT).
\item We model the structural relation between the input image and its low resolution approximation by exploiting the complementary nature of the DWT and SWT, and using particle swarm optimization (PSO).
\item Our approach is the fastest (in terms of CPU time) while producing  satisfactory results.
\end{itemize}
Rest of the paper is organized as follows. The motivation of our proposal is explained in Section \ref{motimation},
our proposed technique is elaborated  in Section \ref{algo} and the comparative results of interpolation for super-resolution using various
techniques including the proposed  approach	 are given in Section \ref{result}. Finally, Section \ref{Conc} concludes the paper.

\section{Motivation}
\label{motimation}
 A super-resolution  system comprises of different sub-modules, namely interpolation, deblurring, artifact removal, denoising, etc. In the system, the pivotal step is interpolation which estimates  values for increasing the resolution of the input image \cite{Park2003}.   Interpolation  is an ill-posed problem as there is no unique solution for the estimation \cite{Dong2013}. Different techniques in literature gather and utilize information relevant to the content of the high resolution image to be generated in different ways.

\begin{table}[]
\caption{Description of important symbols used in the paper}
\begin{tabular}{|l|l|}
\hline
Symbol & Description \\ \hline
 $f(x,y)$      &  Image           \\ \hline
  
  $\varphi$       & Scaling function  \\ \hline
  $\psi$     & Wavelet function  \\ \hline
$R_{kX}$      &   k times the resolution of an image X          \\ \hline
  $\alpha$     & Scaling factor            \\ \hline
  M, N & M is the row size and N is the column size of an image \\ \hline
        ${W_\varphi}$ or $LL $   &  Approximation component after wavelet decomposition            \\ \hline
       ${W^{LH}_\psi}$ or $LH $   &  Vertical details after wavelet decomposition            \\ \hline
    ${W^{HL}_\psi}$ or $HL $   &    Horizontal details after wavelet decomposition          \\ \hline
   ${W^{HH}_\psi}$ or $HH $    & Diagonal details after wavelet decomposition            \\ \hline
  $W^{opt}$ & Optimal weights as model parameters \\ \hline
\end{tabular}
\end{table}

To perform interpolation, in this paper, we consider  human nature regarding anticipation in perception \cite{Bubic2010}. Humans expect and thus predict a certain entity by projecting the knowledge about a contextually similar entity from their past experience. In our interpolation approach, we consider the `past experience' to be the process that estimates the low resolution input from its lower resolution approximation.  The approach  performs the `projection of past experience' by using the aforesaid process to generate the unknown higher resolution image from the low resolution input.  Owing to the said human nature, it would not be inappropriate to assume that a human would do interpolation in a manner  similar to which we consider to design our algorithm.

Now, according to analysis and synthesis of signals using wavelets \cite{Gonzalez2002,Daubechies1990} (\ref{app1}) an image $f(x,y)$ of size $M\times N$ at a higher scale (say $j^{th}$) can be generated through inverse wavelet transform (defined in equation (\ref{eq1})) when we have its approximation  ${W}_\varphi (j-1,m,n)$ (LL), vertical  details ${W^{LH}_\psi} (j-1,m,n)$ (LH), horizontal  details ${W^{HL}_\psi} (j-1,m,n)$ (HL), and diagonal  details ${W^{HH}_\psi} (j-1,m,n)$ (HH) in $(j-1)^{th}$ scale.

\begin{align}
f(x,y) = \frac{1}{{\sqrt {MN} }}\sum\limits_m^{} {\sum\limits_n^{} {{W_\varphi }({j-1},m,n){\varphi _{{j-1},m,n}}(x,y)} } +\nonumber \\ \frac{1}{{\sqrt {MN} }}\sum\limits_{i = LH,HL,HH} {{\sum\limits_m {\sum\limits_n {{W^i}_\psi ({j-1},m,n)} } } } {\psi ^i}_{{j-1},m,n}(x,y)
\label{eq1}
\end{align}
where ${{\varphi _{j-1,m,n}}(x,y)}$ is the scaling function and ${\psi ^{HL}}(j-1,x,y)$, ${\psi ^{LH}}(j-1,x,y)$ and ${\psi ^{HH}}(j-1,x,y)$ are wavelet functions in horizontal, vertical and diagonal direction, respectively, of $(j-1)^{th}$ scale. 
The approximation (LL) and the detail coefficients in horizontal (HL), vertical (LH) and diagonal (HH) directions are generated by wavelet decomposition of the signal $f(x,y)$ as shown in  (\ref{eq1}) using the scaling and wavelet functions.
\begin{equation}
{W_\varphi }({j-1},m,n) = \frac{1}{{\sqrt {MN} }}\sum\limits_{x = 0}^{M - 1} {\sum\limits_{y = 0}^{N - 1} {f(x,y){\varphi _{j-1,m,n}}(x,y)} } 
\label{eq2}
\end{equation}

\begin{equation}
{W^i}_\psi ({j-1},m,n) = \frac{1}{{\sqrt {MN} }}\sum\limits_{x = 0}^{M - 1} {\sum\limits_{y = 0}^{N - 1} {f(x,y){\psi ^i}_{j-1,m,n}(x,y)} } 
\label{eq3}
\end{equation}

If the aforesaid wavelet based analysis and synthesis is considered for interpolation, then,  the input low resolution image can be decomposed into its immediate lower resolution approximation (LL) using (\ref{eq2}) and  high frequency details using (\ref{eq3}). Now, one can assume that only the lower approximation is available for synthesizing back the input image, and not the high frequency details. In such a case, one can try to estimate the unavailable high frequency details 
from the input image's lower approximation. The estimated high frequency details can then be used in (\ref{eq1}) to synthesize an image, which will be an estimate
of the input image.  The estimation of the unavailable  high frequency details should obviously be done to minimize the error between the actual input image and its estimate. Such an estimation process can be considered analogous to modeling the `past experience'. Thus the `projection of past experience' will be analogous to using the process to estimate the input image from its lower approximation for generating (or estimating) the desired high resolution image from the input low resolution image. That is, the input image should be considered as the lower resolution approximation (LL) of the desired high resolution image and the derived estimation model of unavailable  high frequency details should be used to compute the  high frequency details required to generate the desired high resolution image. With the  high frequency details estimated, (\ref{eq1}) can be used for the  aforesaid generation, achieving interpolation.

 The  approach described above is what we employ in our technique and refer to it as the process similarity based interpolation. In the term, process signifies the `past experience' (available knowledge) and similarity allows the `projection of past experience'.
\begin{figure}[!b]
\centering
\includegraphics[width=3.2in]{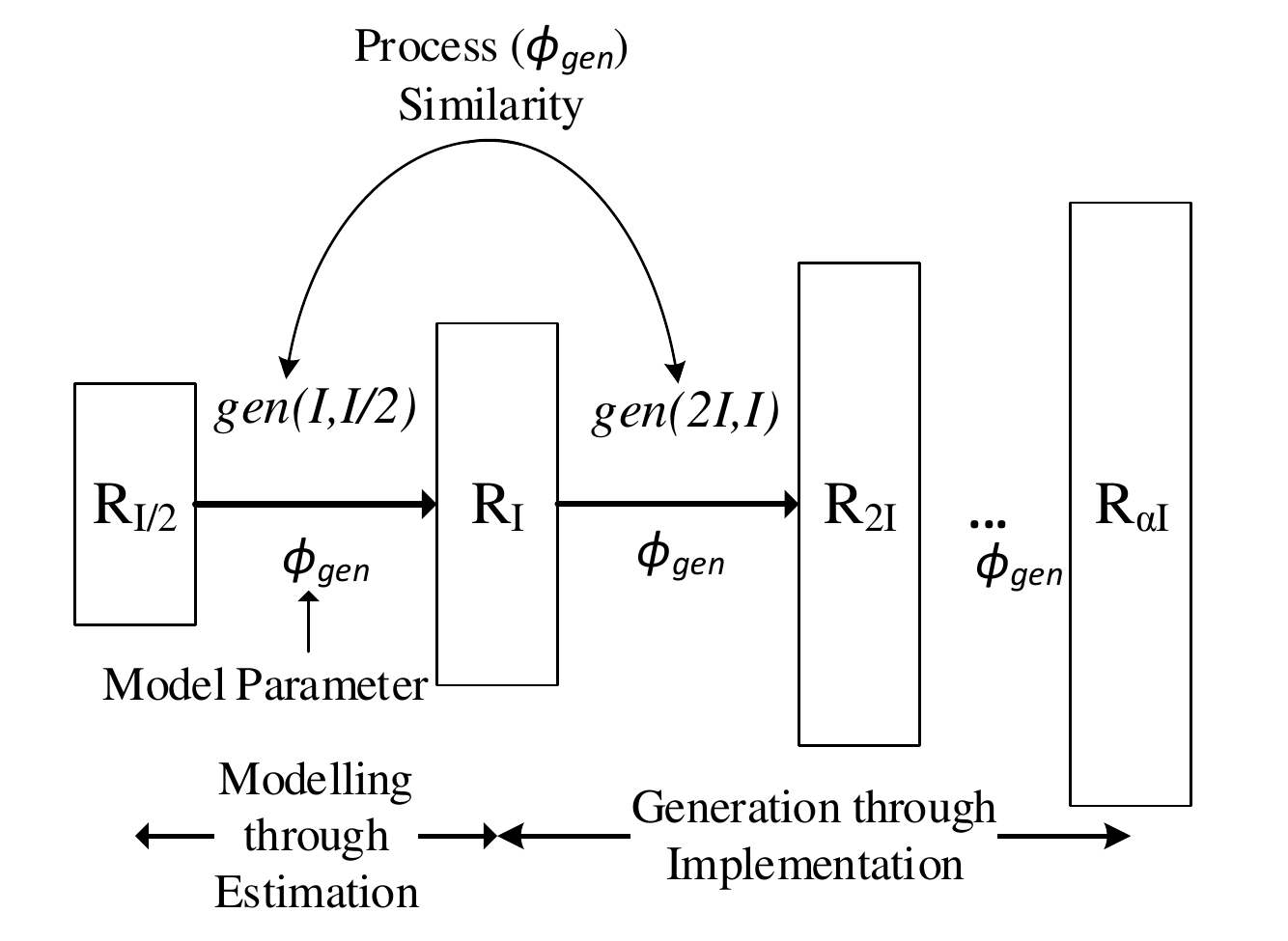}

\caption{Similarity between decompositions at different scales}
\label{fig_pyramid}
\end{figure}
Note that, the use of the process similarity by us is possible due to the resemblance in wavelet decomposition processes at different scales \cite{Carey1999a}. The fundamental idea behind our proposal is that the higher resolution image is generated from a lower resolution image in a similar way  irrespective of the scale.  This can be depicted from the illustration shown in Figure \ref{fig_pyramid}. Let, ${R_I}$ (size: M$\times$N)  denotes the resolution of the input image $I$. So, the output image having  the resolution ${R_{2I}}$ (size: 2M $\times$2N) is assumed to be generated  from the image of  resolution ${R_I}$  in a similar way to the  generation of the image having  resolution ${R_I}$ from the image of  resolution ${R_{I/2}}$ (size: M/2$\times$N/2). It is obvious that the input image at resolution ${R_I}$ and the image at resolution ${R_{I/2}}$ are readily available to optimally model the estimation ($\hat{I}$) of image at ${R_I}$ from the image at ${R_{I/2}}$. Here, we consider input image ($I$) of resolution ${R_I}$ as the ideal one to be estimated from the image of resolution ${R_{I/2}}$  to capture and minimize the loss of information during interpolation. Once the modeling is performed, it can be applied recursively to generate an image of any higher resolution  ${R_{\alpha I}}$  (size: $\alpha$M$\times$$\alpha$N, $\alpha$=$2^l$, $\forall l  \in \mathbb{Z^+}$) from a given input image of resolution ${R_I}$. Note that, by definition, resolution and size of an image may not correspond to each other. As assuming resolution and size to be the same does not affect our analysis, we use them interchangeably for simplicity.

As mentioned earlier, we consider both the discrete wavelet transform (DWT) and the stationary wavelet transform (SWT) in modeling the process similarity. SWT is considered  because of its scale-invariance  property \cite{nason1995stationary}. This allows us to have exact correspondence between spatial locations in an image and its SWT decomposed approximation (LL).  Scale-invariance property does not hold for DWT decomposition, as it involves down-sampling. On the other hand, an application of DWT on an image yields an approximation (LL) with faithful value representation. This is because the basis functions used for DWT decomposition  are not only similar across scales \cite{Carey1999a} but also work at the same translations across scales \cite{Gonzalez2002,Daubechies1990}. This is evident from the  well-known recursion expression relating DWT basis functions at two consecutive scales:
\begin{equation}
\varphi(x-k)=\sqrt 2\sum_nc_n\varphi(2(x-k)-n)
\end{equation}
where $c_n$ is scaling coefficient, and we see that a translation $k$ of the basis function $\varphi(x)$ is related to a translation $2k$ of the down-sampled basis function $\varphi(2x)$, signifying  equal amount of translation.  Such a correspondence does not hold for SWT decomposition.

Note that, although the above discussion of SWT and DWT considers the decomposed approximation (LL), the same holds for the decomposed detail coefficients (HL, LH, and HH) as well. Thus, for SWT decomposition, location correspondence between an image and its decomposed coefficients is accurate. However, the values in the decomposed coefficients might not faithfully represent that in the image. So, we can consider that the location correspondence is our wanted component and the inexact value as the unwanted one in SWT usage. For DWT decomposition, it is the other way around, and hence, we can consider that the location inconsistency is our unwanted component and the value faithfulness as the wanted one.

\FloatBarrier

\section{Proposed process similarity based interpolation}
\label{algo}
Our image interpolation approach uses process similarity (See Section \ref{motimation}) to model the structural relation between scales in the image under consideration. The model is then used to generate the image at the required higher resolution. As elaborated in  Section \ref{Introduction}, across-scale subband self-similarity as in \cite{Singh2014, piao2007image} must not be confused with across-scale process similarity, as the latter gives structural relation whereas the former assumes subband similarity.
   Our proposed algorithm is shown in Figure \ref{fig_algo}. It consists of two modules: modeling and generation. In the modeling stage, the input low resolution image is used to estimate a model capturing the structural relation between scales in  the form of certain optimal weights $W_{1-6}$. In the generation stage, these optimal weights are used to generate the higher resolution image from the input low resolution image. The modules of the proposed technique are elaborated below.

\subsection{Module 1: Modeling}
\label{Modeling}
  \subsubsection{Part 1: High frequency components from discrete (DWT) and shift-invariant (SWT) wavelet decomposition} 
\label{Part1}
As shown in  Figure \ref{fig_algo}, in the proposed approach, DWT decomposes the signal to  half (${R_{I/2}}$) the input resolution ${R_I}$. Four subbands, approximation image LL, vertical details LH, horizontal details HL, and diagonal details HH are obtained with the latter three being the high frequency subbands. Further, the approximation LL is decomposed by both SWT and DWT which will generate subbands at resolutions ${R_{I/2}}$ and ${R_{I/4}}$, respectively. The corresponding high frequency subbands at these two different resolutions are weighted by parameters $W_{1-6}$ and then fused using across-scale addition to generate three components (LH, HL, and HH), which are used in parameter optimization. For across-scale addition, upsampling followed by Gaussian filtering (zero-phase filter) of the lower resolution (${R_{I/4}}$) components is performed before element-wise addition at the higher resolution (${R_{I/2}}$).

The parameter values determined from the optimization (See Section \ref{paramopt})  are tuned to model the structural relation due to process similarity. That is, we consider  components at resolutions ${R_{I/2}}$ and ${R_{I/4}}$ which are used to estimate the image at ${R_I}$, similar to how we will use components at ${R_I}$ and ${R_{I/2}}$ to estimate the desired image at resolution ${R_{2I}}$. Recursive implementation of this operation can generate an image at a desired  high resolution with scaling factor $\alpha$=$2^l$, $\forall l  \in \mathbb{Z^+}$. 

The model parameters (weights $W_{1-6}$) are optimized to generate the LH, HL, and HH components at resolution ${R_{I/2}}$ such that error between the generated image at ${R_{I}}$ and the input image is minimized. Obviously, the image at ${R_{I}}$ is generated using the optimal LH, HL, and HH along with the LL at ${R_{I/2}}$ through IDWT.

\subsubsection{Part 2: Parameter optimization}
\label{paramopt} 
As mentioned in Section \ref{motimation}, our across-scale process similarity based high resolution image generation process uses value faithfulness provided by DWT and location invariance provided by SWT. We use DWT and SWT to get three high frequency subbands that are obtained through across-scale addition.

During across-scale addition, the three components each from DWT and SWT are weighted by model parameters which would represent the structural relation between scales. Among the model  parameters,   the weights $W_{1-3}$ correspond to the three components from DWT and the weights $W_{4-6}$ correspond to those from SWT. To  represent the structural relation between scales  appropriately, these model parameters need to be optimized ($W^{opt}_{1-6}$). To do so, as mentioned in Section \ref{Part1}, we first decompose the input image into lower resolution (${R_{I/2}}$) components and then use the components to optimally get an estimate of the input image. The optimization minimizes the error between the actual input image at resolution ${R_{I}}$ and its estimate.  Particle swarm optimization (PSO)  is a standard and widely used evolutionary optimization algorithm for various applications \cite{zhang2015comprehensive}. PSO is very efficient in reaching an optimal solution utilizing limited memory, and faster computation \cite{kirchmaier2013swarm} with `primitive mathematical operations' \cite{chen2010saliency}. Moreover, the optimization algorithm does not require any prior information about the solution. Global and local explorations of the particles ensure effective and faster optimization. The simple yet robust algorithm can optimize the solution through parallel processing. 
 Therefore, we use  PSO  (as discussed in \ref{app2}) whose details are given in Section \ref{impl}.

Note that the presence of across-scale process similarity allows these optimal weight parameters to represent the structural relation. A lower optimal weight value corresponding to a component from DWT suggests that the error contributed by location correspondence inaccuracy in it is dominating the correctness contributed by its value faithfulness. A higher optimal weight value would signify the opposite. On the other hand, a lower optimal weight value corresponding to a component from SWT suggests that the error contributed by value inexactness in it is dominating the correctness contributed by its accurate location correspondence. Again, a higher optimal weight value would signify the opposite.

Therefore, the optimal weights essentially represent the suitability of each of the six high frequency bands that are used to generate the image with resolution ${R_{I}}$ from the image with resolution ${R_{I/2}}$.  Due to process similarity, as elaborated under the generation  stage, we use these weights to generate an unavailable higher resolution image from the input image at resolution ${R_{I}}$.

\subsection{Module 2: Generation}
In the generation stage,  the model parameters which represent the structural relation, will be used on the input image to generate the image at the desired higher resolution. The approach employs the optimal weights $W^{opt}_{1-6}$ determined in the modeling stage (See Section \ref{Modeling}). The generation block of our algorithm is shown in Figure \ref{fig_algo} within the  dashed (- - -) box. As mentioned earlier, our high resolution image generation process uses DWT and SWT. 

Initially, during the generation, the input image of resolution ${R_{I}}$ is decomposed by DWT and SWT operation. We have four subbands LL, LH, HL, and HH of resolution ${R_{I/2}}$ and ${R_{I}}$ each due to DWT and SWT, respectively. Due to the presence of process similarity, we now use the optimal weights determined in the modeling stage to get the high frequency components to be used in the generation. During fusion by across-scale addition of corresponding high frequency components from DWT and SWT, the six  subbands are multiplied with the corresponding optimal weights. The fusion generates high frequency components LH, HL, and HH at resolution ${R_{I}}$. Then IDWT is applied, where the available input image acts as the approximation LL and the three estimated components LH, HL, and HH as the required high frequency details.

Further, to generate an image of a desired resolution ${R_{\alpha I}}$, where $\alpha$=$2^l$, $l  \in \Bbb Z^{+}$), one  would  require to apply the image generation stage of the proposed interpolation approach  $l$ times recursively.

\section{Experimental Results  and Discussions}
\label{result}
 Experiments are performed to compare the proposed method with six existing methods and the results are reported here. The performance of our approach is compared to that of  edge directed NEDI \cite{Li2001}, nonlocal autoregressive modeling based NARM \cite{Dong2013},  self-exemplar based SR-TSE \cite{Huang2015}, fuzzy predictive composite scheme (FPCS-LAL)\cite{Acharya2017}, and deep learning based EDSR \cite{lim2017enhanced}, SRGAN \cite{ledig2017photo}.  EDSR is a recent technique which performs well for a variety of images and represents the state-of-the-art. The other five approaches are popular in their respective categories, with a couple of them (SRGAN, FPCS-LAL) being recently reported ones. The comparisons here are made both quantitatively and qualitatively in a multifaceted analysis, where the computation time is also considered. Note that, the codes of NARM, SR-TSE, FPCS-LAL, EDSR, and SRGAN were provided by the corresponding authors. We have used the trained models corresponding to the particular decimations for both EDSR and SRGAN  as exactly provided by the corresponding authors.

  \subsection{Implementation details:}
\label{impl}
 Our algorithm is based on wavelet decomposition, and hence it is  required to choose a wavelet function (/scaling function). As our approach analyzes the representation of an image in its approximation and detail high frequency components to perform interpolation, we choose two different wavelet functions for the experiments, which differ substantially in representing local regularity of a signal. We consider Daubechies wavelet with support 4 (db2) and discrete approximation of Meyer filter (support is 102), where the latter excels in approximating signals with low local regularity/smoothness. The proposed method using Daubechies wavelet is referred here as  `proposed (Daub)' and that using discrete Meyer wavelet as `proposed (Meyer)'. Note that, the proposed (Daub) and proposed (Meyer) follow exactly the same approach (See Figure \ref{fig_algo}) except for the wavelet used in decomposition and reconstruction.

Our algorithm also uses PSO where we need to provide initialization and stopping criteria. We empirically observed that stopping with the maximum number of iterations as 15 after initializing the population with 15 randomly generated particles in 6-dimensional space ($W_{1-6}$, $\forall$ $W_i$ $\in$ [0 1]) works efficiently to produce at least near-optimal solution. Our PSO is designed for early termination. It stops if there is no change in the fitness value of the best solution in consecutive runs after the number of iterations is greater than 7. The minimum number of iterations is considered as 7 to allow sufficient exploration of the solution space  $ S = \{ \boldmath {X} \in \mathbb{R}{^6}|\forall {x} \in [0,1]\} $. PSNR is considered as the fitness/objective function whose maximization results in minimization of error between the desired image ${I}$ and the generated image ${I_W^'}$. This optimization function is as  follows  
\begin{equation}
{W^{opt}} = \underset{W}{\operatorname{argmax}}{PSNR ({I}, {I_W^'})}
\end{equation}
 The source code of our approach  can be downloaded from here\footnote{\href{https://drive.google.com/open?id=1jHw9r6gvT7AC8EnmozBV4z4U8AvFnAOe}{https://drive.google.com/open?id=1jHw9r6gvT7AC8EnmozBV4z4U8AvFnAOe}}.\subsection{Computation speed}
\label{Com_Time}
The increasing developments in display technology for electronic gadgets and  content delivery through data streaming demands faster interpolation for immediate display of content without compromising much on quality. Such real-life applications require an algorithm to get high resolution, good quality output as fast as possible. Therefore, as far as interpolation is concerned in today's world, the computation time of an algorithm is still an important factor.

\begin{table} [!h]
\caption{Computation time comparison between different interpolation algorithm for scaling factor 2 (Best result in bold)		}
\begin{center}
\label{time_scale2}
\scalebox{1}{%

\begin{tabular}{|l|l|l|l|l|l|l|}
\hline
\multirow{2}{*}{\begin{tabular}[c]{@{}l@{}}Input Low\\  Resolution\end{tabular}} & \multicolumn{6}{c|}{Technique (speed in seconds}                                                                                                                         \\ \cline{2-7} 
                                                                                 & NEDI & NARM & SR-TSE & FPCS-LAL & \begin{tabular}[c]{@{}l@{}}Proposed\\ (Daub)\end{tabular} & \begin{tabular}[c]{@{}l@{}}Proposed\\ (Meyer)\end{tabular} \\ \hline
$64\times 64 $                                                                          &  0.593  & 13.81
     & 7.57  & \textbf{0.38}   & \textbf{\makecell{0.38 \\ $\pm$ 0.022}}  & \makecell{0.93 \\ $\pm$ 0.075}                                  \\ \hline
$128 \times 128 $                                                                       &    2.68  &  64.16
    & 34.57  & 4.23   & \textbf{\makecell{0.54
 \\ $\pm$ 0.014}}  &  \makecell{1.40
 \\ $\pm$ 0.119}                                                    \\ \hline
$256 \times 256 $ &   11.56   &   290.37
   & 187.72  & 152.65   & \textbf{\makecell{0.93
 \\ $\pm$ 0.093}} &  \makecell{2.71 \\ $\pm$ 0.218}                                                      \\ \hline

\end{tabular}}

\end{center}
\end{table}

\begin{table}[!h]
\caption{Computation time  (CPU time) comparison  between different interpolation algorithm for scaling factor 4 (Best result in bold)}
\begin{center}
\label{time_scale4}
\scalebox{1}{%

\begin{tabular}{|l|l|l|l|l|l|l|}
\hline
\multirow{2}{*}{\begin{tabular}[c]{@{}l@{}}Input Low\\  Resolution\end{tabular}} & \multicolumn{5}{c|}{Technique (speed in seconds)}                                                                                                                         \\ \cline{2-6} 
                                                                                 & NEDI & SR-TSE & FPCS-LAL & \begin{tabular}[c]{@{}l@{}}Proposed\\ (Daub)\end{tabular} & \begin{tabular}[c]{@{}l@{}}Proposed\\ (Meyer)\end{tabular} \\ \hline
$64\times 64 $                                                                          &  3.22  
     & 22.00 & 4.36  &\textbf{\makecell{1.23 \\ $\pm$ 0.108}}  & \makecell{2.74 \\ $\pm$ 0.208}                                  \\ \hline
$128 \times 128 $                                                                       &    13.70  
    & 113.18 & 148.65   & \textbf{\makecell{1.78 \\ $\pm$ 0.036}}  &  \makecell{4.72
 \\ $\pm$ 0.370}                                                    \\ \hline
$256 \times 256 $ &   54.65   
   & 534.77 & 3072   & \textbf{\makecell{5.79
 \\ $\pm$ 0.201}} &  \makecell{12.75 \\ $\pm$ 1.0}                                                      \\ \hline

	\end{tabular}}

\end{center}
\end{table}

So, in this paper, we consider the computation time  as an essential  parameter to judge the efficiency of an algorithm. We use an Intel\textsuperscript \textregistered  i5(3.30 GHz) machine with 16 GB RAM and use MATLAB\textsuperscript \textregistered (version 2016a) for this purpose.  Our target is to perform fast interpolation which generates a high resolution image exploiting process similarity at low resolution without introducing any visual artifact. To calculate the computation speed, we have considered the 16, $ 512 \times 512$ images from USC SIPI Miscellaneous gray scale image database\footnote{\href{http://sipi.usc.edu/database/}{http://sipi.usc.edu/database/}}, and down-sampled them to generate images of $ 64 \times 64 $, $ 128 \times 128 $ and $ 256 \times 256 $ sizes. We consider gray scale images of above mentioned three  sizes as we perform an analysis of computation speed variations with the change in 2D matrix size (image size). The average computation time (CPU time) of all the algorithms over 16 images of a particular size are presented in Table \ref{time_scale2} and \ref{time_scale4} for scaling factors 2 and 4, respectively. As the proposed approach is based on heuristic optimization, the  computation time may slightly vary due to early termination and random initialization. So, for our algorithm, we have considered 5 runs for each image to present the average computation time (CPU time) (over 16$\times$5 runs) along with the standard deviation, which is negligible. For NARM, we have shown the computation time (CPU time) only for scaling factor 2 as the code\footnote{\href{http://www4.comp.polyu.edu.hk/~cslzhang/NARM.htm}{http://www4.comp.polyu.edu.hk/~cslzhang/NARM.htm}}  provided does not have an implementation of factor 4.  Deep learning based algorithms such as EDSR, and SRGAN  require a considerable number of low resolution (LR)-high resolution (HR) image pairs for learning, which (CPU time) consumes  considerable amount of time in graphical processing units (GPUs) \cite{cruz2018single}. Moreover, the  implementations of EDSR and SRGAN by their authors (trained networks) also require GPUs to generate the output. As a fair comparison of computation times between approaches implemented in CPU and GPU is not possible, we do not include the same.

From Table \ref{time_scale2}, it is evident that FPCS-LAL and  proposed (Daub) algorithm have the least computation time (CPU time) of 0.38 seconds for an input image of $ 64 \times 64 $ yielding an output of size $ 128 \times 128 $. If the input size increases, our  proposed (Daub) becomes substantially faster compared to the other algorithms compared here. Our  proposed (Meyer) algorithm is  slightly slower than the proposed (Daub) algorithm, but much faster when compared with other  interpolation  techniques presented in the table. Table \ref{time_scale4}, where the scaling factor is 4, shows that for an input image of size $ 64 \times 64 $ to $ 256 \times 256 $, our proposed algorithms are much faster than all the existing algorithms compared here. So, our proposed  approach is most suited among all the algorithms here when  time is an important factor of application.  

 It is to be noted that the faster speed achieved by our algorithms is not by
utilizing large memory space. This is because memory is occupied only by the
multi-scale decompositions up to two levels, and the input and output images. In 
addition, a little memory is required \cite{Poli2007} for the PSO particles and parameters.

\subsection{Subjective  Evaluation:}

Computation time (CPU time) comparison above shows that our proposed algorithms are considerably faster than the other algorithms  and the  difference increases in favor of our algorithms with the increase of the input image size and scaling factor.

With the above observation, it becomes important to analyze and compare the interpolation performance. Before we go into quantitative evaluation, where decimation/down-sampling of images from databases would be required, we  present qualitative evaluation here. For qualitative analysis, we consider images from databases  as input images for interpolation, and generate twice and four times larger images by the various interpolation approaches. For this experiment, we consider USC SIPI Miscellaneous image database (24 gray images and 16 color images). While the entire set of results is shared online\footnote[1]{\href{https://drive.google.com/file/d/1-X-ENfISl1tSlOOJsWAR5FxIHn-OSl0s/view?usp=sharing/}{https://drive.google.com/file/d/1-X-ENfISl1tSlOOJsWAR5FxIHn-OSl0s/view?usp=sharing/}}, we consider a few examples here to summarize the generic observations on interpolation by the various approaches.
     
\begin{figure}[!h]
\centering
\begin{subfigure}[b]{.10\linewidth}
\includegraphics[width=\linewidth]{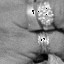}
\caption{Original}\label{fig:mouse}
\end{subfigure}
\hspace{88pt}
\begin{subfigure}[b]{.36\linewidth}
\includegraphics[width=\linewidth]{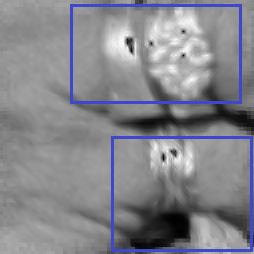}
\caption{NEDI}\label{fig:mouse}
\end{subfigure}

\begin{subfigure}[b]{.36\linewidth}
\includegraphics[width=\linewidth]{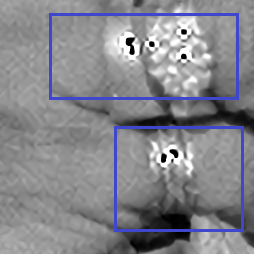}
\caption{SR-TSE}\label{fig:mouse}
\end{subfigure}
\begin{subfigure}[b]{.36\linewidth}
\includegraphics[width=\linewidth]{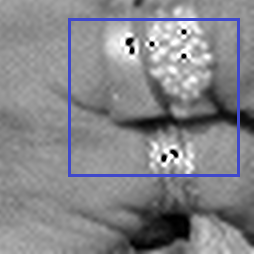}
\caption{FPCS-LAL}\label{fig:mouse}
\end{subfigure}
\begin{subfigure}[b]{.36\linewidth}
\includegraphics[width=\linewidth]{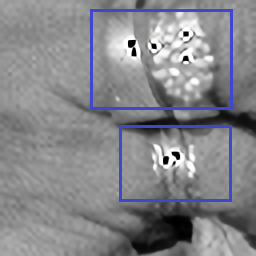}
\caption{EDSR}\label{fig:mouse}
\end{subfigure}
\begin{subfigure}[b]{.36\linewidth}
\includegraphics[width=\linewidth]{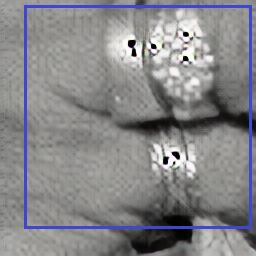}
\caption{SRGAN}\label{fig:mouse}
\end{subfigure}
\begin{subfigure}[b]{.36\linewidth}
\includegraphics[width=\linewidth]{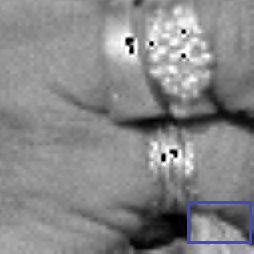}
\caption{Proposed(Daub)}\label{fig:sf4GrayDB}
\end{subfigure}
\begin{subfigure}[b]{.36\linewidth}
\includegraphics[width=\linewidth]{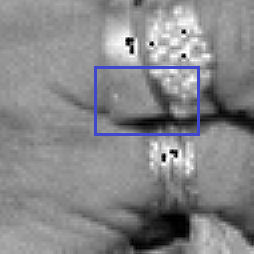}
\caption{Proposed(Meyer)}\label{fig:mouse}
\end{subfigure}
\caption{ A cropped version of  gray scale Man image for scaling factor of 4}
\label{fig:sf_gray4}
\end{figure}

\begin{figure}[!h]
\centering
\begin{subfigure}[b]{.10\linewidth}
\includegraphics[width=\linewidth]{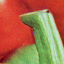}
\caption{Original}\label{fig:mouse}
\end{subfigure}
\hspace{88pt}
\begin{subfigure}[b]{.36\linewidth}
\includegraphics[width=\linewidth]{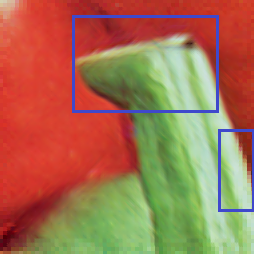}
\caption{NEDI}\label{fig:mouse}
\end{subfigure}

\begin{subfigure}[b]{.36\linewidth}
\includegraphics[width=\linewidth]{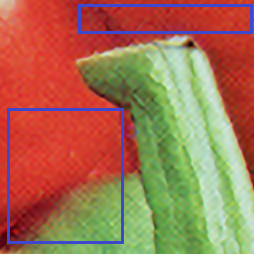}
\caption{SR-TSE}\label{fig:mouse}
\end{subfigure}
\begin{subfigure}[b]{.36\linewidth}
\includegraphics[width=\linewidth]{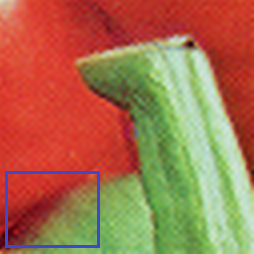}
\caption{FPCS-LAL}\label{fig:mouse}
\end{subfigure}
\begin{subfigure}[b]{.36\linewidth}
\includegraphics[width=\linewidth]{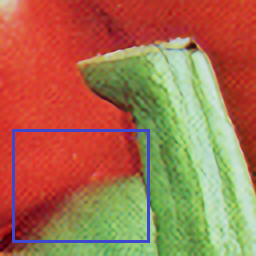}
\caption{EDSR}\label{fig:mouse}
\end{subfigure}
\begin{subfigure}[b]{.36\linewidth}
\includegraphics[width=\linewidth]{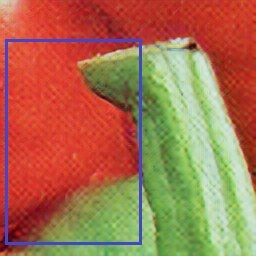}
\caption{SRGAN}\label{fig:mouse}
\end{subfigure}
\begin{subfigure}[b]{.36\linewidth}
\includegraphics[width=\linewidth]{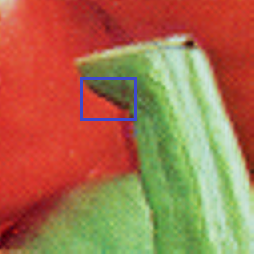}
\caption{Proposed(Daub)}\label{fig:sf4colorDB}
\end{subfigure}
\begin{subfigure}[b]{.36\linewidth}
\includegraphics[width=\linewidth]{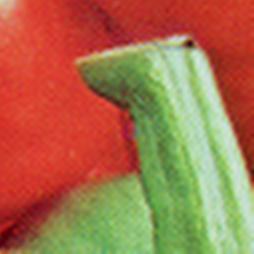}
\caption{Proposed(Meyer)}\label{fig:mouse}
\end{subfigure}
\caption{ A cropped version of  Pepper image for scaling factor of 4}
\label{fig:sf_color4}
\end{figure}

In Figure \ref{fig:sf_gray4}, we show the subjective comparison on a gray scale image considering the scaling factor to be 4. The cropped portion of  the whole image is shown here for better visualization\footnote[2]{Visualization is appropriate when looked  at 100\% on a computer screen. The observations made later may not be visible in print.}. We see that NEDI's output is smoother than all the others, which results in blurring and distortions especially at regions having changing gray values. SR-TSE, FPCS-LAL, and EDSR produce much sharper outputs than NEDI but are prone to ringing artifacts at regions of drastic change and appearance of the non-existent pattern in homogeneous regions. SRGAN produces artifacts which look like fixed pattern noise.  Our approach produces negligible amount of the artifacts mentioned above, but are prone to some amount of jaggies at the boundary regions. As can be seen, the overall impression from the results is that the proposed algorithms  perform better than the others.
In Figure \ref{fig:sf_color4}, we show another such (scaling factor 4) subjective comparison now on a color image. We again see here the artifacts mentioned above in the results by NEDI, SR-TSE, FPCS-LAL, EDSR, SRGAN and our proposed algorithms.  Further, we can  see that the artifacts in the results of the proposed algorithms  is less pronounced compared to that in the others.

In Figures \ref{fig:sf_gray4} and \ref{fig:sf_color4}, although both the proposed algorithms  do better than the others and also contain some amount of jaggies artifact, there are differences in their results. We have observed in our experiments that the jaggies introduced by the proposed algorithms are mainly due to aliasing, but with different characteristics. As can be seen, in Figure \ref{fig:sf_gray4}, the effect of aliasing by the proposed (Meyer) algorithm is concentrated at the mid-frequencies (like texture regions). On the other hand, aliasing by the  proposed (Daub) algorithm is spread across all frequencies in small amount and is pronounced at high frequencies (sharp edges, see Figures \ref{fig:sf4GrayDB} and \ref{fig:sf4colorDB}).

\begin{figure}[!h]
\centering
\begin{subfigure}[b]{.16\linewidth}
\includegraphics[width=\linewidth]{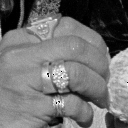}
\caption{Original}\label{fig:mouse}
\end{subfigure}
\hspace{52pt}
\begin{subfigure}[b]{.32\linewidth}
\includegraphics[width=\linewidth]{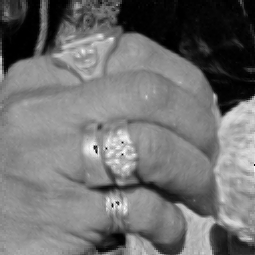}
\caption{NEDI}\label{fig:mouse}
\end{subfigure}
\begin{subfigure}[b]{.32\linewidth}
\includegraphics[width=\linewidth]{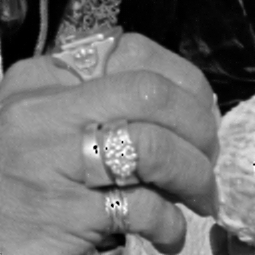}
\caption{NARM}\label{fig:mouse}
\end{subfigure}
\begin{subfigure}[b]{.32\linewidth}
\includegraphics[width=\linewidth]{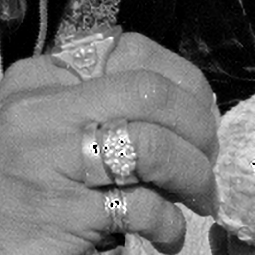}
\caption{SR-TSE}\label{fig:mouse}
\end{subfigure}
\begin{subfigure}[b]{.32\linewidth}
\includegraphics[width=\linewidth]{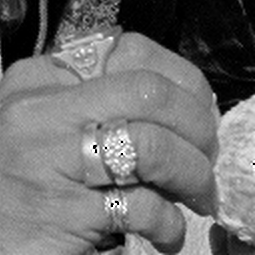}
\caption{FPCS-LAL}\label{fig:mouse}
\end{subfigure}
\begin{subfigure}[b]{.32\linewidth}
\includegraphics[width=\linewidth]{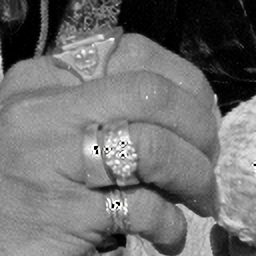}
\caption{EDSR}\label{fig:mouse}
\end{subfigure}
\begin{subfigure}[b]{.32\linewidth}
\includegraphics[width=\linewidth]{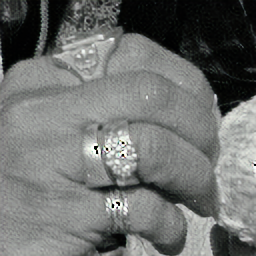}
\caption{SRGAN}\label{fig:mouse}
\end{subfigure}\begin{subfigure}[b]{.32\linewidth}
\includegraphics[width=\linewidth]{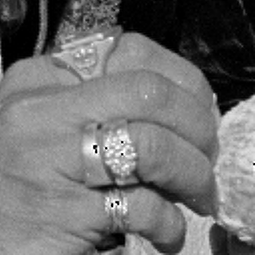}
\caption{Proposed(Daub)}\label{fig:mouse}
\end{subfigure}
\begin{subfigure}[b]{.32\linewidth}
\includegraphics[width=\linewidth]{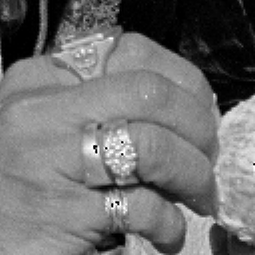}
\caption{Proposed(Meyer)}\label{fig:mouse}
\end{subfigure}

\caption{ A cropped version of  gray scale Man image for scaling factor of 2}
\label{fig:sf_gray2}
\end{figure}

\begin{figure}[!h]
\centering
\begin{subfigure}[b]{.16\linewidth}
\includegraphics[width=\linewidth]{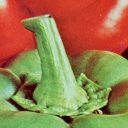}
\subcaption{Original}\label{fig:mouse}
\end{subfigure}
\hspace{52pt}
\begin{subfigure}[b]{.32\linewidth}
\includegraphics[width=\linewidth]{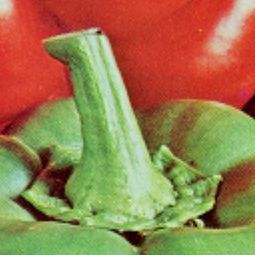}
\caption{NEDI}\label{fig:mouse}
\end{subfigure}
\begin{subfigure}[b]{.32\linewidth}
\includegraphics[width=\linewidth]{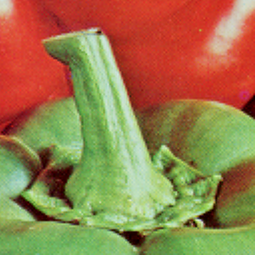}
\caption{NARM}\label{fig:mouse}
\end{subfigure}
\begin{subfigure}[b]{.32\linewidth}
\includegraphics[width=\linewidth]{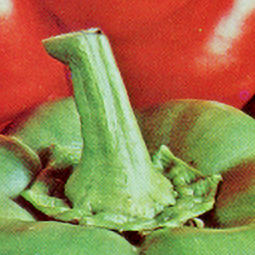}
\caption{SR-TSE}\label{fig:mouse}
\end{subfigure}
\begin{subfigure}[b]{.32\linewidth}
\includegraphics[width=\linewidth]{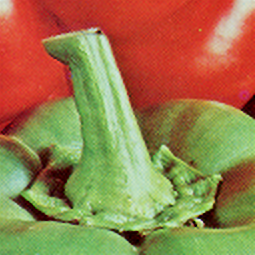}
\caption{FPCS-LAL}\label{fig:mouse}
\end{subfigure}
\begin{subfigure}[b]{.32\linewidth}
\includegraphics[width=\linewidth]{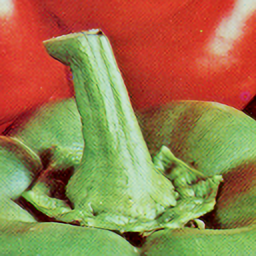}
\caption{EDSR}\label{fig:mouse}
\end{subfigure}
\begin{subfigure}[b]{.32\linewidth}
\includegraphics[width=\linewidth]{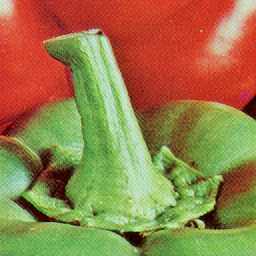}
\caption{SRGAN}\label{fig:mouse}
\end{subfigure}
\begin{subfigure}[b]{.32\linewidth}
\includegraphics[width=\linewidth]{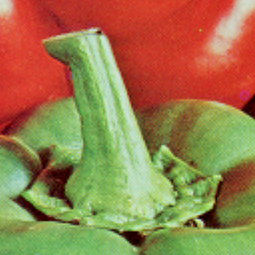}
\caption{Proposed(Daub)}\label{fig:mouse}
\end{subfigure}
\begin{subfigure}[b]{.32\linewidth}
\includegraphics[width=\linewidth]{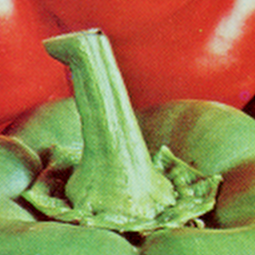}
\caption{Proposed(Meyer)}\label{fig:mouse}
\end{subfigure}
\caption{ A cropped version of  color Pepper image for scaling factor of 2}
\label{fig:sf_color2}
\end{figure}

Further, in Figures \ref{fig:sf_gray2} and \ref{fig:sf_color2}, we show subjective comparison considering scaling factor 2 on gray and color images, respectively. Some of the observations similar to Figures \ref{fig:sf_gray4} and \ref{fig:sf_color4} such as NEDI's smoother output and appearance of  the non-existent pattern in outputs of SR-TSE, FPCS-LAL, SRGAN, and EDSR are evident upon close observation. NARM, like NEDI, also   produces a smooth output. However, upon quick observation it is hard to find a substantial difference between all the results. In such a situation, it is obvious that one should prefer an algorithm with the least computation time (CPU time), which is, our  proposed (Daub) algorithm.

	 Now,  let us consider a situation where  it is possible for a user to choose an algorithm for an image. The following may be done based on the   subjective evaluation of interpolation performance. In such a scenario, the  proposed (Daub) algorithm should be chosen for images having a good amount of contents with changing pixel values (mid frequency) like textures. Whereas, the proposed (Meyer) algorithm should be employed if the changes in an image are mostly drastic (high frequency) and it has a good amount of homogeneous regions (low frequency). However, if for any input image, it is not possible for a user to choose an algorithm among proposed (Daub) and proposed (Meyer) for generating the output, then proposed (Daub) algorithm should automatically be chosen because of

\begin{itemize}
\item[1.] The lower computation time (CPU time).
\item[2.] The fact that natural images contain very less high frequency contents than low and mid frequency ones.
\end{itemize}

\subsection{Quantitative Evaluation:}
\label{quant}
Unlike subjective evaluation, to evaluate the methods quantitatively, we need a known target/reference based on which the generated high resolution image can be assessed. Therefore, we generate the low resolution input images considering the actual images in databases as the desired high resolution images.  

As stated in \cite{Dong2013}, low resolution images to be used as the inputs are usually obtained by decimation (smoothing followed by down-sampling) of those images that are considered as the desired high resolution images. As stated  in \cite{Demirel2011,Tian2011}, wavelet approximation filter is used for the smoothing, as said in \cite{ledig2017photo}, Gaussian filter is used for smoothing,  as stated in \cite{Li2001, Mallat2010} and \cite{Dong2011} only down-sampling without smoothing is considered. The smoothing in the decimation can be achieved using different filters with different properties. However, an interpolation algorithm might be particularly suited to a kind of smoothing used to produce the low resolution input for quantitative evaluation. This could happen when the interpolation uses operators/filters consistent with a particular kind of decimation.

Therefore, we choose to treat the kind of smoothing used in the decimation as a parameter. To evaluate an algorithm's quantitative performance corresponding to an image, we calculate the PSNR, SSIM and FSIM values for five types of inputs obtained through the following operations
 \begin{itemize}
  \item  Bicubic approximation(Bicubic)
  \item  Wavelet filter approximation (Daubechies)
  \item  Wavelet filter approximation (D-Meyer)
  \item  Gaussian filtering followed by down-sampling (Gaussian)
  \item  Down-sampling without any smoothing (Sub-sampling)
  \end{itemize}
For quantitative comparisons, experiments are carried out using standard USC SIPI Miscellaneous image database (24 gray images and 16 color images), BSD image database (100 color images), SET 5 (5 color images) and SET 14 (14 color and gray scale images) \cite{Huang2015}. These images offer different types of challenges for an interpolation algorithm to deal with. 
As PSO based optimization in our algorithm can yield slightly different results in different runs, like in Section \ref{Com_Time}, we consider the average performance over 5 runs of our algorithm on each image. These average performances are obtained for all the images in a database, and the average over all these images are reported. The standard deviation quantifying variation in average performance over all images in a database during the 5 runs is found to be in the order of $10^{-3}$ dB (PSNR), and hence it is not reported individually. For all the different cases, the quantitative results are listed in tables with the best performing result in bold.

  \begin{table} [!h] 
\caption{PSNR(dB) comparison of gray scale images from USC SIPI Miscellaneous database	for scaling factor 2	}
\begin{center}
\label{Tab_PSNR_gray2}
\scalebox{0.9}{%

\begin{tabular}{|l|l|l|l|l|l|l|l|l|}
\hline
\multirow{2}{*}{\begin{tabular}[c]{@{}l@{}}Input Low\\  Resolution\end{tabular}} & \multicolumn{8}{c|}{Technique}                                                                                                                         \\ \cline{2-9} 
                                                                                 & NEDI & NARM & SR-TSE & FPCS-LAL & EDSR & SRGAN  & \begin{tabular}[c]{@{}l@{}}Proposed\\ (Daub)\end{tabular} & \begin{tabular}[c]{@{}l@{}}Proposed\\ (Meyer)\end{tabular} \\ \hline
Bicubic                                                                          &   27.60     &   27.78
   & 31.24  & 30.41   &  \textbf{32.07}  &  30.83 & 29.27                                                   & 28.09                                                      \\ \hline
Daubechies                                                                       &    28.25  &  29.14
    & 29.15  & 29.31   & 29.22  & 24.14   & \textbf{30.24}                                                   & 29.39                                                      \\ \hline
D-Meyer                                                                          &   28.95   &   30.00
   & 27.05 & 27.35   & 26.96  & 26.31   & 29.09                                                   & \textbf{30.68}                                                      \\ \hline
Gaussian                                                                         &   28.55   &   29.12
   & 27.94  & 27.91    & 27.99  & 27.45  & 29.11                                                   & \textbf{29.93}                                                      \\ \hline
Sub-sampling                                                                    &    28.06 &   \textbf{28.96}
   & 24.70  & 25.30    & 24.40  & 28.40  & 27.46                                                   & 28.02                                                      \\ \hline
Average                                                                          &   28.28   &   29.00   & 28.02  & 28.06   & 28.13  & 27.43   & {29.03}                                                   & \textbf{29.22}                                                      \\ \hline

\end{tabular}}

\end{center}
\end{table}

   \begin{table}[!h]
\caption{SSIM comparison of gray scale images from USC SIPI Miscellaneous database	for scaling factor 2	}
\begin{center}
\label{Tab_SSIM_gray2}
\scalebox{0.9}{%

\begin{tabular}{|l|l|l|l|l|l|l|l|l|}
\hline
\multirow{2}{*}{\begin{tabular}[c]{@{}l@{}}Input Low\\  Resolution\end{tabular}} & \multicolumn{8}{c|}{Technique}                                                                                                                         \\ \cline{2-9} 
                                                                                 & NEDI & NARM & SR-TSE & FPCS-LAL & EDSR & SRGAN & \begin{tabular}[c]{@{}l@{}}Proposed\\ (Daub)\end{tabular} & \begin{tabular}[c]{@{}l@{}}Proposed\\ (Meyer)\end{tabular} \\ \hline
Bicubic                                                                          & 0.7972 &   0.8013
                        & 0.8867 & 0.8714  &\textbf{0.8995}  & 0.8859  & 0.8427                         & 0.8216                                                     \\ \hline
Daubechies                                                                       & 0.8280 &   0.8433
                        & 0.8543 & 0.7607  & 0.8619	& 0.7076 & \textbf{0.8786}                                                  & 0.8587                                                     \\ \hline
D-Meyer                                                                          & 0.8337 &         0.8540
                  & 0.8066 & 0.8166 & 0.8108	& 0.7834  & 0.8526                                                  & \textbf{0.8765}                                                     \\ \hline
Gaussian                                                                         & 0.8204 &       0.8271
                    & 0.8238 & 0.8208 &  0.8293	& 0.8133  & 0.8494                                                  & \textbf{0.8578}                                                     \\ \hline
Sub-sampling                                                                    & 0.8340 &       \textbf{0.8495}
                   & 0.7330 & 0.7581 &  0.7316	& 0.8407  & 0.8244                     & 0.8218                                                     \\ \hline
Average                                                                          &  0.8223 &   0.8350 &   0.8208  &     0.8055   &  0.8266	& 0.8062
   &   \textbf{0.8495} &                                                        {0.8472} \\ \hline

\end{tabular}}

	\end{center}
\end{table}
\FloatBarrier

 \begin{table} [!h]
\caption{FSIM comparison of gray scale images from USC SIPI Miscellaneous database	for scaling factor 2	}
\begin{center}
\label{Tab_FSIM_gray2}
\scalebox{0.9}{%

\begin{tabular}{|l|l|l|l|l|l|l|l|l|}
\hline
\multirow{2}{*}{\begin{tabular}[c]{@{}l@{}}Input Low\\  Resolution\end{tabular}} & \multicolumn{8}{c|}{Technique}                                                                                                                         \\ \cline{2-9} 
                                                                                 & NEDI & NARM & SR-TSE & FPCS-LAL & EDSR & SRGAN  & \begin{tabular}[c]{@{}l@{}}Proposed\\ (Daub)\end{tabular} & \begin{tabular}[c]{@{}l@{}}Proposed\\ (Meyer)\end{tabular} \\ \hline
Bicubic                                                                          & 0.9342 &    0.9365
                       & 0.9694 & 0.9618  & \textbf {0.9747} 	& 0.973  & 0.9479                   & 0.9381                                                     \\ \hline
Daubechies                                                                       & 0.9451 &    0.9520
                       & 0.9699 & 0.9658   &  0.9711	& 0.9075        & \textbf{0.9724   } & 0.9503                                                     \\ \hline
D-Meyer                                                                          & 0.9527 &     \textbf{0.9616}
                      & 0.9286 & 0.9313   &  0.9296	& 0.9267   & 0.9558                   & \textbf{0.9616}                                                     \\ \hline
Gaussian                                                                         & 0.9437 &   0.9500
                        & 0.9352 & 0.9337  &  0.9365	& 0.9338
   & 0.9576                                                  & \textbf{0.9578}                                                     \\ \hline
Sub-sampling                                                       & 0.9426 &   0.9491                  & 0.9093 & 0.9103   &  0.9092	& \textbf{0.9649}
  & 0.9335                                                  & 0.9361                                                     \\ \hline
Average                                                                          &  0.9437      & {0.9499} & 0.9424      & 0.9405     &  0.9442	& 0.9412
       &                                                        \textbf{0.9534} &                                                           0.9487 \\ \hline
 
\end{tabular}}

\end{center}
\end{table} 
\FloatBarrier
  Tables \ref{Tab_PSNR_gray2}, \ref{Tab_SSIM_gray2} and \ref{Tab_FSIM_gray2} show comparative results considering images from USC SIPI Miscellaneous gray scale image database in terms of PSNR, SSIM, and FSIM, respectively, for the scaling factor of 2. Table \ref{Tab_PSNR_gray2} shows that among all the algorithms compared on the basis of PSNR, EDSR performs the best for input generated by bicubic approximation,  proposed (Daub) algorithm  for input generated by Daubechies wavelet approximation, the proposed (Meyer) algorithm  for input generated by Meyer wavelet approximation and Gaussian filtering followed by down-sampling, and NARM for input generated by only down-sampling.
  
But on an average over all types of  inputs, the performance of our proposed approach is superior. PSNR value of proposed (Meyer) and  proposed (Daub) algorithms are respectively 0.22 dB and 0.03 dB higher than the next best, NARM. Tables \ref{Tab_SSIM_gray2} and \ref{Tab_FSIM_gray2} respectively considering SSIM and FSIM also show that different algorithms perform the best for different input types, with the proposed approach being the best for Daubechies and Meyer wavelet approximations, and Gaussian filtering followed by down-sampling. On an average over all types of  inputs, our  proposed (Daub) algorithm yields slightly better results.

\begin{table}[!h]
\caption{PSNR(dB) comparison of color images from USC SIPI Miscellaneous database	for scaling factor 2	}
\begin{center}
\label{Tab_PSNR_USCColor2}
\scalebox{0.9}{%

\begin{tabular}{|l|l|l|l|l|l|l|l|l|}
\hline
\multirow{2}{*}{\begin{tabular}[c]{@{}l@{}}Input Low\\  Resolution\end{tabular}} & \multicolumn{8}{c|}{Technique}                                                                                                                         \\ \cline{2-9} 
                                                                                 & NEDI & NARM & SR-TSE & FPCS-LAL & EDSR & SRGAN  & \begin{tabular}[c]{@{}l@{}}Proposed\\ (Daub)\end{tabular} & \begin{tabular}[c]{@{}l@{}}Proposed\\ (Meyer)\end{tabular} \\ \hline
Bicubic                                                                          &  30.60  & 31.13
     & 36.48  & 34.92 & \textbf{37.82}	& 35.59
    & 33.18                                                   & 31.12                                                      \\ \hline
Daubechies                                                                       &    31.23  &  33.25
    & 33.85  & 33.75   &  33.95	& 32.69
 & \textbf{34.90}                                                   & 33.22                                                      \\ \hline
D-Meyer                                                                          &   32.49   &   35.12
   & 30.09  & 30.26  &  29.75	& 29.11
  & 32.59                                                   & \textbf{35.47}                                                      \\ \hline
Gaussian                                                                         &  32.11    &  33.74
    & 31.24  & 31.00   &  31.18	& 30.64
 & 32.67                                                   & \textbf{34.37}                                                      \\ \hline
Sub-sampling                                                                    &  32.00    &  \textbf{34.56}
    & 28.34  & 28.76  & 27.94 & 27.42 & 31.64                                                   & 33.11                                                      \\ \hline
Average                                                                          &   31.66   &    \textbf{33.56}  & 32.00     & 31.74 &  32.13	& 31.09
   & 33.00                                                   & {33.46}                                                      \\ \hline

\end{tabular}}

\end{center}
\end{table}

\begin{table}[!h]
\caption{PSNR(dB) comparison of color images from BSD database	for scaling factor 2	}
\begin{center}
\label{Tab_PSNR_BSD2}
\scalebox{0.9}{%

\begin{tabular}{|l|l|l|l|l|l|l|l|l|}
\hline
\multirow{2}{*}{\begin{tabular}[c]{@{}l@{}}Input Low\\  Resolution\end{tabular}} & \multicolumn{8}{c|}{Technique}                                                                                                                         \\ \cline{2-9} 
                                                                                 & NEDI & NARM & SR-TSE & FPCS-LAL & EDSR & SRGAN  & \begin{tabular}[c]{@{}l@{}}Proposed\\ (Daub)\end{tabular} & \begin{tabular}[c]{@{}l@{}}Proposed\\ (Meyer)\end{tabular} \\ \hline
Bicubic                                                                          &  27.63    & 27.84
     & 31.12  & 30.55   & \textbf{ 32.17}	& 31.22
       & 29.53                                                   & 28.16                                                      \\ \hline
Daubechies                                                                       &  28.55    &  29.37
    & 29.12  & 29.18   &  29.07	& 28.53
      & \textbf{30.10}                                                   & 29.31                                                      \\ \hline
D-Meyer                                                                          &   29.22   &  30.29
    & 26.95  & 27.24   &  26.7	& 26.46
      & 28.82                                                   & \textbf{30.70}                                                      \\ \hline
Gaussian                                                                         &  28.74    &  29.31
    & 27.92  & 27.91  & 27.96	& 27.74
       & 28.85                                                   & \textbf{29.82}                                                      \\ \hline
Sub-sampling                                                                    &   28.78   &    \textbf{29.39}
  & 24.59  & 25.16   &  23.99	& 24.01
    & 27.51                                                   & 28.00                                                      \\ \hline
Average                                                                          &  28.54    &  \textbf{29.24}    & 27.94  & 28.00 &  27.98	& 27.59
   & 28.97                                                   & {29.20}      
                                                \\ \hline
 
\end{tabular}}

\end{center}
\end{table}

\begin{table}[!h]
\caption{PSNR(dB) comparison of color images from SET 5 database for  scaling factor 2	}
\begin{center}
\label{Tab_PSNR_Set5_2}
\scalebox{0.9}{%

\begin{tabular}{|l|l|l|l|l|l|l|l|l|}
\hline
\multirow{2}{*}{\begin{tabular}[c]{@{}l@{}}Input Low\\  Resolution\end{tabular}} & \multicolumn{8}{c|}{Technique}                                                                                                                         \\ \cline{2-9} 
                                                                                 & NEDI & NARM & SR-TSE & FPCS-LAL & EDSR & SRGAN  & \begin{tabular}[c]{@{}l@{}}Proposed\\ (Daub)\end{tabular} & \begin{tabular}[c]{@{}l@{}}Proposed\\ (Meyer)\end{tabular} \\ \hline
Bicubic      & 30.33 & 30.49 & 36.46  & 35.19    & \textbf{37.97} & 35.54 & 33.48 & 30.65 \\ \hline
Daubechies   & 28.95 & 32.73 & 33.41  & 33.13    & 33.13 & 31.71 & \textbf{34.35} & 32.72 \\ \hline
D-Meyer      & 32.8  & 34.93  & 29.63  & 29.75    & 29.37 & 28.78 & 31.98 & \textbf{35.44} \\ \hline
Gaussian     & 32.15 & 33.27 & 30.65  & 30.56    & 30.69 & 30.35 & 31.93 & \textbf{33.97} \\ \hline
Sub-sampling & 32.58 & \textbf{34.79} & 28.05  & 28.35    & 27.44 & 26.95 & 31.00    & 33.07 \\ \hline
Average      & 31.36 & \textbf{33.24} & 31.64  & 31.40     & 31.72 & 30.66 & 32.55 & {33.17}\\ \hline
 
\end{tabular}}

\end{center}
\end{table}

\begin{table}[!h]
\caption{PSNR(dB) comparison of gray scale and color images from SET 14 database for  scaling factor 2	}
\begin{center}
\label{Tab_PSNR_Set14_2}
\scalebox{0.9}{%

\begin{tabular}{|l|l|l|l|l|l|l|l|l|}
\hline
\multirow{2}{*}{\begin{tabular}[c]{@{}l@{}}Input Low\\  Resolution\end{tabular}} & \multicolumn{8}{c|}{Technique}                                                                                                                         \\ \cline{2-9} 
                                                                                 & NEDI & NARM & SR-TSE & FPCS-LAL & EDSR & SRGAN  & \begin{tabular}[c]{@{}l@{}}Proposed\\ (Daub)\end{tabular} & \begin{tabular}[c]{@{}l@{}}Proposed\\ (Meyer)\end{tabular} \\ \hline

Bicubic      & 27.59 & 27.94 & 32.29  & 31.28    & \textbf{33.30}  & 31.90  & 29.94 & 28.03 \\ \hline
Daubechies   & 28.43 & 29.81 & 28.47  & 29.89    & 28.06 & 28.91 & \textbf{30.76} & 29.71 \\ \hline
D-Meyer      & 29.42 & 31.18 & 26.99  & 27.29    & 26.63 & 26.31 & 26.24 & \textbf{31.52} \\ \hline
Gaussian     & 28.95 & 30.00    & 28.01  & 27.96    & 28.03 & 27.74 & 29.09 & \textbf{30.42} \\ \hline
Sub-sampling & 28.94 & \textbf{30.51} & 25.05  & 25.56    & 24.19 & 24.14 & 27.96 & 29.09 \\ \hline
Average      & 28.67 & \textbf{29.89} & 28.16  & 28.40     & 28.04 & 27.80  & 28.80  & {29.75}\\ \hline
 
\end{tabular}}

\end{center}
\end{table}

\begin{table}[!h]
\caption{SSIM comparison of color images from USC SIPI Miscellaneous database	for scaling factor 2	}
\begin{center}
\label{Tab_SSIM_USC Color2}
\scalebox{0.9}{%

\begin{tabular}{|l|l|l|l|l|l|l|l|l|}
\hline
\multirow{2}{*}{\begin{tabular}[c]{@{}l@{}}Input Low\\  Resolution\end{tabular}} & \multicolumn{8}{c|}{Technique}                                                                                                                         \\ \cline{2-9} 
                                                                                 & NEDI & NARM & SR-TSE & FPCS-LAL & EDSR & SRGAN  & \begin{tabular}[c]{@{}l@{}}Proposed\\ (Daub)\end{tabular} & \begin{tabular}[c]{@{}l@{}}Proposed\\ (Meyer)\end{tabular} \\ \hline
Bicubic                                                                          &   0.8803
   &   0.8847
   & 0.9359  & 0.9274 &  \textbf{0.9431}	& 0.9247
  & 0.9124                                                & 0.8927                                                   \\ \hline
Daubechies                                                                       &   0.9001
   &  0.9119
    & 0.9195 & 0.9220 &  0.9257	& 0.8988
  & \textbf{0.9278}                                                & 0.9157                                                   \\ \hline
D-Meyer                                                                          &   0.9066
   &   0.9192
   & 0.8817 & 0.8872  &  0.8817	& 0.8521
  & 0.9082                                                & \textbf{0.9273}                                                   \\ \hline
Gaussian                                                                         &  0.8975
    &   0.9045
   & 0.8954 & 0.8929 &  0.8986	& 0.8818
  & 0.9079                                                & \textbf{0.9185}                                                   \\ \hline
Sub-sampling                                                                    &   0.9062 &  \textbf{0.9188}
    & 0.8351 & 0.8505 &  0.8327	& 0.7912
  & 0.8931                                                & 0.8955                                                   \\ \hline
Average                                                                          &  0.8981   & 0.9078     & 0.8935 & 0.8960 & 0.8964	&0.8697
  & {0.9099}  & \textbf{0.9100} \\ \hline
 
\end{tabular}}

\end{center}
\end{table}

\begin{table}[!h]
\caption{SSIM comparison of color images from BSD database		for scaling factor 2}
\begin{center}
\label{Tab_SSIM_BSD2}
\scalebox{0.9}{%

\begin{tabular}{|l|l|l|l|l|l|l|l|l|}
\hline
\multirow{2}{*}{\begin{tabular}[c]{@{}l@{}}Input Low\\  Resolution\end{tabular}} & \multicolumn{8}{c|}{Technique}                                                                                                                         \\ \cline{2-9} 
                                                                                 & NEDI & NARM & SR-TSE & FPCS-LAL & EDSR & SRGAN  & \begin{tabular}[c]{@{}l@{}}Proposed\\ (Daub)\end{tabular} & \begin{tabular}[c]{@{}l@{}}Proposed\\ (Meyer)\end{tabular} \\ \hline
Bicubic                                                                          &    0.7873
  &    0.7938
  & 0.8856 & 0.8740 & \textbf{0.899} 	& 0.8843
   & 0.8509 & 0.8264                                                    \\ \hline
Daubechies                                                                       &    0.8217
  &    0.8421
  & 0.8556 & 0.8574 & 0.8634 &	0.8429
  & \textbf{0.8716} & 0.8544 \\ \hline
D-Meyer                                                                          &    0.8331
  &   0.8562
   & 0.8064 & 0.8135 & 0.8058	& 0.7838 & 0.8449 & \textbf{0.8784}                                            \\ \hline
Gaussian                                                                         &    0.8112
  &   0.8218
   & 0.8172  & 0.8136 & 0.8224	& 0.8109
    &  0.8351                                                & \textbf{0.8518}                                                   \\ \hline
Sub-sampling                                                                    &  0.8353
    &     \textbf{0.8511}
 & 0.7323 & 0.7558 & 0.7235	& 0.7063
  & 0.8257                                                & 0.8141                                                   \\ \hline
Average                                                                          &   0.8177   &   {0.8330}   & 0.8194 & 0.8228 & 0.8228	& 0.8057
  & \textbf{0.8457} & \textbf{0.8457}                                                      \\ \hline

\end{tabular}}

\end{center}
\end{table}

\begin{table}[!h]
\caption{SSIM comparison of color images from SET 5 database for   scaling factor 2	}
\begin{center}
\label{Tab_SSIM_Set5_2}
\scalebox{0.9}{%

\begin{tabular}{|l|l|l|l|l|l|l|l|l|}
\hline
\multirow{2}{*}{\begin{tabular}[c]{@{}l@{}}Input Low\\  Resolution\end{tabular}} & \multicolumn{8}{c|}{Technique}                                                                                                                         \\ \cline{2-9} 
                                                                                 & NEDI & NARM & SR-TSE & FPCS-LAL & EDSR & SRGAN  & \begin{tabular}[c]{@{}l@{}}Proposed\\ (Daub)\end{tabular} & \begin{tabular}[c]{@{}l@{}}Proposed\\ (Meyer)\end{tabular} \\ \hline

Bicubic      & 0.8935  & 0.8956 & 0.9532 & 0.944    & \textbf{0.9594} & 0.9398 & 0.9306 & 0.9041 \\ \hline
Daubechies   & 0.8869 & 0.9248 & 0.9333 & 0.93     & 0.9343 & 0.9045 & \textbf{ 0.9388} & 0.9283 \\ \hline
D-Meyer      & 0.9212 & 0.9346 & 0.8908 & 0.893    & 0.8887 & 0.8548 & 0.9179 & \textbf{0.9423} \\ \hline
Gaussian     & 0.9115  & 0.9184 & 0.9073 & 0.9043   & 0.9097 & 0.8916 & 0.9173 & \textbf{0.9332} \\ \hline
Sub-sampling & 0.9224  & \textbf{0.9386} & 0.8479 & 0.8591   & 0.8376 & 0.8005 & 0.9049 & 0.9148\\ \hline
Average      & 0.9071  & {0.9224} & 0.9065 & 0.9061   & 0.9059 & 0.8782 & 0.9219 & \textbf{0.9245} \\ \hline
 
\end{tabular}}

\end{center}
\end{table}

\begin{table}[!h]
\caption{SSIM comparison of gray scale and color images from SET 14 database for  scaling factor 2	}
\begin{center}
\label{Tab_SSIM_Set14_2}
\scalebox{0.9}{%

\begin{tabular}{|l|l|l|l|l|l|l|l|l|}
\hline
\multirow{2}{*}{\begin{tabular}[c]{@{}l@{}}Input Low\\  Resolution\end{tabular}} & \multicolumn{8}{c|}{Technique}                                                                                                                         \\ \cline{2-9} 
                                                                                 & NEDI & NARM & SR-TSE & FPCS-LAL & EDSR & SRGAN  & \begin{tabular}[c]{@{}l@{}}Proposed\\ (Daub)\end{tabular} & \begin{tabular}[c]{@{}l@{}}Proposed\\ (Meyer)\end{tabular} \\ \hline
Bicubic      & 0.8135 & 0.8229 & 0.9042 & 0.8928   & \textbf{0.9153} & 0.8955 & 0.8722 & 0.8359 \\ \hline
Daubechies   & 0.8459 & 0.8671 & 0.8533 & 0.8780    & 0.8442 & 0.8552 & \textbf{0.8896} & 0.8741 \\ \hline
D-Meyer      & 0.8572 & 0.8806 & 0.8257 & 0.8324   & 0.8218 & 0.7950  & 0.7231 & \textbf{0.8950}  \\ \hline
Gaussian     & 0.8393 & 0.8527 & 0.8397 & 0.8361   & 0.844  & 0.8274 & 0.8560  & \textbf{0.8744} \\ \hline
Sub-sampling & 0.8573 & \textbf{0.8783} & 0.7586 & 0.7810    & 0.7431 & 0.7162 & 0.8431 & 0.8497 \\ \hline
Average      & 0.8426 & {0.8603} & 0.8363 & 0.8441   & 0.8337 & 0.8179 & 0.8368 & \textbf{0.8658} \\ \hline
 
\end{tabular}}

\end{center}
\end{table}

\begin{table}[!h]
\caption{FSIM comparison of color images from USC SIPI Miscellaneous database	for scaling factor 2	}
\begin{center}
\label{Tab_FSIM_USCColor2}
\scalebox{0.9}{%

\begin{tabular}{|l|l|l|l|l|l|l|l|l|}
\hline
\multirow{2}{*}{\begin{tabular}[c]{@{}l@{}}Input Low\\  Resolution\end{tabular}} & \multicolumn{8}{c|}{Technique}                                                                                                                         \\ \cline{2-9} 
                                                                                 & NEDI & NARM & SR-TSE & FPCS-LAL & EDSR & SRGAN  & \begin{tabular}[c]{@{}l@{}}Proposed\\ (Daub)\end{tabular} & \begin{tabular}[c]{@{}l@{}}Proposed\\ (Meyer)\end{tabular} \\ \hline
Bicubic                                                                          &  0.9407
    &   0.9422
   & 0.9814 & 0.9727 &\textbf{0.9866}  & 0.9807
 & 0.9631                                                & 0.9466                                                   \\ \hline
Daubechies                                                                       &  0.9570
    &  0.9627
    & 0.9719 & 0.9706 & 0.9729	& 0.9656
 & \textbf{0.9765}                                                & 0.9634                                                    \\ \hline
D-Meyer                                                                          &   0.9658
   &     0.9731
 & 0.9398 & 0.9418 & 0.9399	& 0.9329
 & 0.9596                                                & \textbf{0.9753}                                                   \\ \hline
Gaussian                                                                         &   0.9578
   &   0.9622
   & 0.9487 & 0.9463 & 0.9499	& 0.9461
 & 0.9608                                                & \textbf{0.9704}                                                   \\ \hline
Sub-sampling                                                                    &   0.9640
   &    \textbf{0.9704}
  & 0.9225 & 0.9266 & 0.9234	& 0.9149
 & 0.9516                                                & 0.9587                                                   \\ \hline
Average                                                                          &   0.9570   &   {0.9621}   & 0.9528 & 0.9516 & 0.9546 & 0.9481
 & 0.9517 & \textbf{0.9629}                                                     \\ \hline

\end{tabular}}

\end{center}
\end{table}

\begin{table}[!h]
\caption{FSIM comparison of color images from BSD database		for scaling factor 2}
\begin{center}
\label{Tab_FSIM_BSD2}
\scalebox{0.9}{%

\begin{tabular}{|l|l|l|l|l|l|l|l|l|}
\hline
\multirow{2}{*}{\begin{tabular}[c]{@{}l@{}}Input Low\\  Resolution\end{tabular}} & \multicolumn{8}{c|}{Technique}                                                                                                                         \\ \cline{2-9} 
                                                                                 & NEDI & NARM & SR-TSE & FPCS-LAL & EDSR & SRGAN  & \begin{tabular}[c]{@{}l@{}}Proposed\\ (Daub)\end{tabular} & \begin{tabular}[c]{@{}l@{}}Proposed\\ (Meyer)\end{tabular} \\ \hline
Bicubic                                                                          &    0.8682
  &   0.8731
   & 0.9400 & 0.9283 & \textbf{0.9503} 	& 0.9441
    & 0.9114                                                 & 0.9063                                                   \\ \hline
Daubechies                                                                       &    0.8966
  &    0.9080
  & 0.9188 & 0.9192 & 0.9224	& 0.9148
   & \textbf{0.9288} & 0.9169 \\ \hline
D-Meyer                                                                          &  0.9051
    &  0.9189
    & 0.8820 & 0.8888 & 0.8815	& 0.8745
     & 0.9092                                                & \textbf{0.9289}                                                  \\ \hline
Gaussian                                                                         &  0.8815
    &    0.8890
  & 0.8959 & 0.8925 & 0.8967	& 0.8973
   & 0.9033 & \textbf{0.9143} \\ \hline
Sub-sampling                                                                    &   0.9161
   &   \textbf{0.9264}
   & 0.8490 & 0.8634 & 0.8458	& 0.8407
     & 0.9040 & 0.8885 \\ \hline
Average                                                                          &  0.8935    &  0.9031    & 0.8971 & 0.8984 & 0.8993	& 0.8943
 &\textbf{0.9113} & {0.9110} \\ \hline

\end{tabular}}

\end{center}
\end{table}

\begin{table}[!h]
\caption{FSIM comparison of color images from SET 5 database for scaling factor 2	}
\begin{center}
\label{Tab_FSIM_Set5_2}
\scalebox{0.9}{%

\begin{tabular}{|l|l|l|l|l|l|l|l|l|}
\hline
\multirow{2}{*}{\begin{tabular}[c]{@{}l@{}}Input Low\\  Resolution\end{tabular}} & \multicolumn{8}{c|}{Technique}                                                                                                                         \\ \cline{2-9} 
                                                                                 & NEDI & NARM & SR-TSE & FPCS-LAL & EDSR & SRGAN  & \begin{tabular}[c]{@{}l@{}}Proposed\\ (Daub)\end{tabular} & \begin{tabular}[c]{@{}l@{}}Proposed\\ (Meyer)\end{tabular} \\ \hline
Bicubic      & 0.9257 & 0.9262 & 0.9748 & 0.9604   & \textbf{0.9822} & 0.9739 & 0.9499 & 0.9307 \\ \hline
Daubechies   & 0.9344 & 0.9503 & 0.9554 & 0.9488   & 0.9572 & 0.9432 & \textbf{0.9599} & 0.9483 \\ \hline
D-Meyer      & 0.9498 & \textbf{0.9586} & 0.9173 & 0.9166   & 0.9164 & 0.9029 & 0.9408 & 0.9574\\ \hline
Gaussian     & 0.9407 & 0.9454 & 0.9370  & 0.9320    & 0.9381 & 0.9347 & 0.9438 & \textbf{0.9548} \\ \hline
Sub-sampling & 0.9577 & \textbf{0.9677} & 0.8979 & 0.9013   & 0.8954 & 0.8827 & 0.9390  & 0.9422 \\ \hline
Average      & 0.9417 & \textbf{0.9496} & 0.9456 & 0.9318   & 0.9379 & 0.9275 & {0.9467} & {0.9467} \\ \hline
 
\end{tabular}}

\end{center}
\end{table}

\begin{table}[!h]
\caption{FSIM comparison of gray scale and color images from SET 14 database for  scaling factor 2	}
\begin{center}
\label{Tab_FSIM_Set14_2}
\scalebox{0.9}{%

\begin{tabular}{|l|l|l|l|l|l|l|l|l|}
\hline
\multirow{2}{*}{\begin{tabular}[c]{@{}l@{}}Input Low\\  Resolution\end{tabular}} & \multicolumn{8}{c|}{Technique}                                                                                                                         \\ \cline{2-9} 
                                                                                 & NEDI & NARM & SR-TSE & FPCS-LAL & EDSR & SRGAN  & \begin{tabular}[c]{@{}l@{}}Proposed\\ (Daub)\end{tabular} & \begin{tabular}[c]{@{}l@{}}Proposed\\ (Meyer)\end{tabular} \\ \hline
Bicubic      & 0.9174 & 0.9212 & 0.973  & 0.9637   & \textbf{0.9783} & 0.9727 & 0.9501 & 0.9286 \\ \hline
Daubechies   & 0.9406 & 0.9489 & 0.9562 & 0.9602   & 0.9325 & 0.9530  & \textbf{0.9684} & 0.9523 \\ \hline
D-Meyer      & 0.9520  & 0.9628 & 0.9224 & 0.9243   & 0.9209 & 0.9168 & 0.8756 & \textbf{0.9684} \\ \hline
Gaussian     & 0.9382 & 0.9454 & 0.9302 & 0.9278   & 0.9312 & 0.9295 & 0.9454 & \textbf{0.9578} \\ \hline
Sub-sampling & \textbf{0.9489} & 0.9580  & 0.8990  & 0.9067   & 0.8953 & 0.8926 & 0.9354 & 0.9459 \\ \hline
Average      & 0.9394 & {0.9472} & 0.9362 & 0.9366   & 0.9316 & 0.9329 & 0.9350  & \textbf{0.9506} \\ \hline
 
\end{tabular}}

\end{center}
\end{table}

The computationally intensive algorithms proposed in \cite{Yang2010, Dong2013},  generate high resolution color images,  by applying their proposed interpolation technique on the luminance component and a simple interpolation technique like the cubic \cite{keys1981cubic} method to generate chromatic component\footnote{This is evident from the codes provided as well}. As our approach is computationally fast, we leverage this advantage and apply our interpolation to all the channels.  The benefits of our approach are thus reflected in all the three channels.

 Tables \ref{Tab_PSNR_USCColor2} - \ref{Tab_FSIM_Set14_2} show comparative results in terms of PSNR, SSIM,  and FSIM  considering scaling factor of 2 for the  images from the USC SIPI Miscellaneous color and BSD, SET 5, and SET 14  databases. Here, we see the technique performing the best for different kinds of inputs is similar, that is, EDSR performs the best for input generated by bicubic approximation,  proposed (Daub) algorithm for input generated by Daubechies wavelet approximation, proposed (Meyer) algorithm for input generated by Meyer wavelet approximation (except FSIM in SET 5) and NARM for input generated by only down-sampling.  For Gaussian, the best performing algorithm is our proposed (Meyer). On an average over all types of inputs,  the performance of our algorithm is comparable with the best result  (that of NARM) in terms of PSNR. However, our algorithms' average performance is the best in terms of SSIM for all databases and also in terms of FSIM except for SET 5 database.

Tables \ref{Tab_PSNR_GRAY4} - \ref{Tab_FSIM_Set14_4} show the comparative results of PSNR, SSIM and FSIM similar to that of Tables \ref{Tab_PSNR_gray2} - \ref{Tab_FSIM_Set14_2}, except that the scaling factor is 4 instead of 2. Here, we find that different techniques perform the best for different kinds of inputs. On an average over all types of  inputs,  the  performance of our algorithm is the best for all the databases except for USC SIPI Miscellaneous gray database in terms of PSNR. But in terms of SSIM and FSIM, our techniques' performance is comparable with the best which is one of the algorithms among SR-TSR, EDSR, and SRGAN.

The above quantitative analysis shows that on an average performance over all types of inputs considered here, our proposed algorithms in terms of PSNR, SSIM and FSIM perform either the best or near to the best considering different databases containing color and gray scale images, and different scaling factors. Among the proposed algorithms, we do not find enough evidence here to choose one over the other. But one can choose the  proposed (Daub) algorithm based on the observations made during qualitative analysis.

 So, through the computation time (CPU time), qualitative and quantitative analyses, we find that the proposed approach not only consistently produces the best or near-best results but also does so in the least time (CPU time). Moreover, the  lower computation time (CPU time) becomes more significant with the increase in the input image size and scaling factor. 
 
  Note that, our proposed approach suffers from a few drawbacks due to the use of heuristic
optimization, PSO. Due to the stochastic behavior of PSO, our approach can 
yield slightly different results in different runs. Although the performance for
quantitative evaluation shows that there can be a very slight variation from the average values, the algorithms may not produce the best result possible by them
in a particular run. Moreover, the computation time may slightly vary in each
run due to the early termination. In addition, as mentioned already, although 
our approach produces negligible amount of artifacts than the other algorithms,
it is prone to appearance of some amount of jaggies at the boundary regions.

\subsection{Auxiliary analysis}

We provide a couple of additional analysis of the proposed approach. The first analysis is regarding the fact that the proposed approach is designed to generate images considering any scaling factor of the form $2^l$, $\forall l  \in \mathbb{Z^+}$. When one desires a scaling factor $\alpha$, which cannot be expressed as $2^l$, we suggest that a value of $l$ is considered such that $2^{l-1}<\alpha<2^l$. Then, the interpolated image of larger size can be decimated by $\alpha/2^l$ using any simple and fast symmetric spline \cite{spath1995two} approximation.

\begin{figure} [!h]
\centering
\begin{subfigure}[b]{.44\linewidth}
\includegraphics[width=\linewidth]{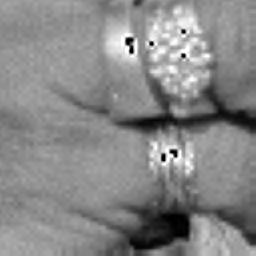}
\caption{$\times4$}\label{fig:gray4}
\end{subfigure}
\begin{subfigure}[b]{.33\linewidth}
\includegraphics[width=\linewidth]{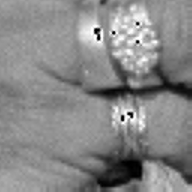}
\caption{$\times3$}\label{fig:gray3}
\end{subfigure}
\begin{subfigure}[b]{.44\linewidth}
\includegraphics[width=\linewidth]{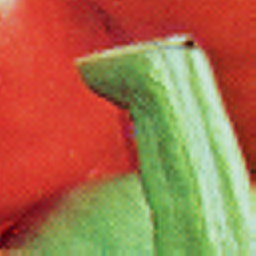}
\caption{$\times4$}\label{fig:color4}
\end{subfigure}
\begin{subfigure}[b]{.33\linewidth}
\includegraphics[width=\linewidth]{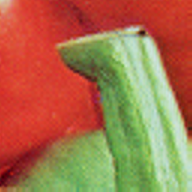}
\caption{$\times3$}\label{fig:color3}
\end{subfigure}

\caption{Subjective Performance of our  proposed (Daub) for Color and Gray Scale Image}
\label{fig:sf3}
\end{figure}
\FloatBarrier

We show the results obtained  by performing such an interpolation of scaling factor 3 using the  proposed (Daub) algorithm in Figure \ref{fig:sf3}. As we see for both the gray and color images, no additional artifacts have been introduced in the images of Figures   \ref{fig:gray3} and \ref{fig:color3} due to the decimation using bicubic approximation applied on Figures \ref{fig:gray4} and \ref{fig:color4}, respectively.

The second analysis here is regarding further improvement in computation time (CPU time) of the proposed approach. In this analysis, we show that by compromising a little in the interpolation performance of the proposed approach, results can be obtained much faster. Consider  Table \ref{Time_Aux}, which shows computation times for  proposed (Daub) algorithm similar to those discussed in Tables \ref{time_scale2} and \ref{time_scale4}. However, the computation times in Table \ref{Time_Aux} have been obtained by altering the  proposed (Daub) algorithm's PSO based optimization, where instead of zero change in fitness value during iteration, a change $\le5\times 10^{-3}$ dB is checked for termination. As can be seen, comparing the  computation times in Table \ref{Time_Aux}, and Tables \ref{time_scale2} and \ref{time_scale4}, there is about $35\%$ average improvement in the case where a scaling factor is 2 and about $40\%$ average improvement when a scaling factor is 4. With the alteration in the  proposed (Daub) algorithm, it was observed that the average interpolation performance over all images in the USC SIPI Miscellaneous gray scale database  was reduced by  only around  $10^{-2}$ dB  (in terms of PSNR) when input images were obtained by Daubechies wavelet filter approximation.

The final analysis provided here is on performance efficiency. Considering the USC SIPI Miscellaneous gray image database, Table~\ref{fom} lists the average performance for the scaling factor 2 of different approaches along with the average time (CPU time) taken to achieve them. Doing so, it sheds some light on the rate of quantitative performance   with respect to computational expenses. As can be seen, the proposed approach is the most efficient one,  achieving the best average performance in the least average time.

\subsection{Domain-specific Interpolation}
In this section, we aim to analyse the performance and applicability of our proposed approach for domain specific interpolation. For this, we consider interpolation of human face and text images.
 
\begin{figure}
\centering
\begin{subfigure}[b]{.14\linewidth}
\includegraphics[width=\linewidth]{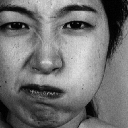}
\caption{Original}\label{fig:mouse}
\end{subfigure}
\hspace{52pt}
\begin{subfigure}[b]{.14\linewidth}
\includegraphics[width=\linewidth]{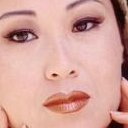}
\caption{Original}\label{fig:mouse}
\end{subfigure}\\
\begin{subfigure}[b]{.27\linewidth}
\includegraphics[width=\linewidth]{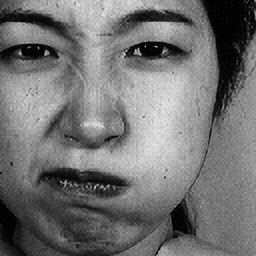}
\caption{EDSR}\label{fig:mouse}
\end{subfigure}
\begin{subfigure}[b]{.27\linewidth}
\includegraphics[width=\linewidth]{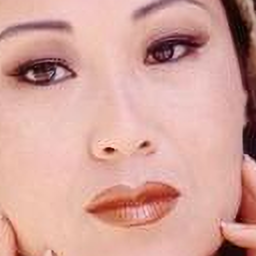}
\caption{EDSR}\label{fig:mouse}
\end{subfigure}\\
\begin{subfigure}[b]{.27\linewidth}
\includegraphics[width=\linewidth]{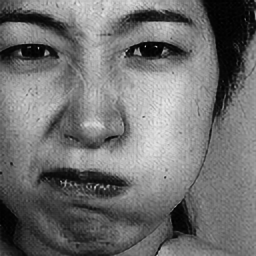}
\caption{SRGAN}\label{fig:mouse}
\end{subfigure}
\begin{subfigure}[b]{.27\linewidth}
\includegraphics[width=\linewidth]{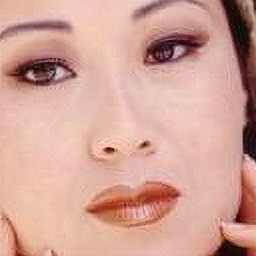}
\caption{SRGAN}\label{fig:mouse}
\end{subfigure}\\
\begin{subfigure}[b]{.27\linewidth}
\includegraphics[width=\linewidth]{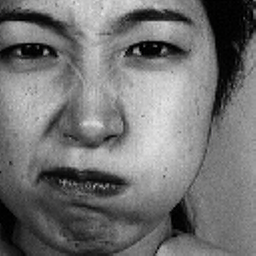}
\caption{Proposed(Daub)}\label{fig:mouse}
\end{subfigure}
\begin{subfigure}[b]{.27\linewidth}
\includegraphics[width=\linewidth]{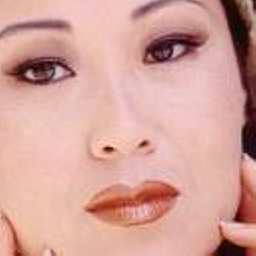}
\caption{Proposed(Daub)}\label{fig:mouse}
\end{subfigure}\\
\begin{subfigure}[b]{.27\linewidth}
\includegraphics[width=\linewidth]{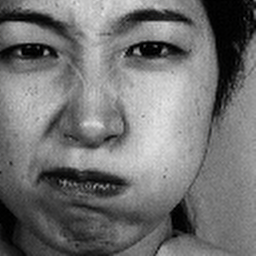}
\caption{Proposed(Meyer)}\label{fig:mouse}
\end{subfigure}
\begin{subfigure}[b]{.27\linewidth}
\includegraphics[width=\linewidth]{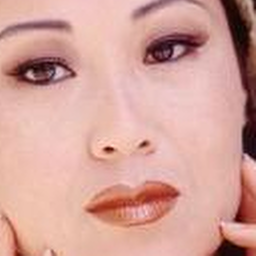}
\caption{Proposed(Meyer)}\label{fig:mouse}
\end{subfigure}

\caption{ Cropped version of  a gray scale face image (left column) and a color face image (right column)  for scaling factor of 2}
\label{Sub_Face22}
\end{figure}
\urlstyle{same}
Human face images are images where a wide range of frequencies exist just like many other natural images. As we have shown earlier that our proposed algorithm performs well for natural images (five databases), we expect it to do well in face images as well. For the experiment on face images, we consider the standard gray scale face images from Jaffe\footnote{\url{http://www.kasrl.org/jaffe.html}} (213 images) database and the three color face images from SET5 database.

\begin{figure}
\centering
\begin{subfigure}[b]{.07\linewidth}
\includegraphics[width=\linewidth]{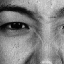}
\caption{Original}\label{fig:mouse}
\end{subfigure}
\hspace{52pt}
\begin{subfigure}[b]{.07\linewidth}
\includegraphics[width=\linewidth]{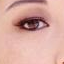}
\caption{Original}\label{fig:mouse}
\end{subfigure}\\
\begin{subfigure}[b]{.27\linewidth}
\includegraphics[width=\linewidth]{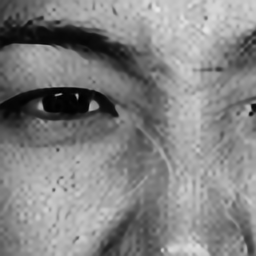}
\caption{EDSR}\label{fig:mouse}
\end{subfigure}
\begin{subfigure}[b]{.27\linewidth}
\includegraphics[width=\linewidth]{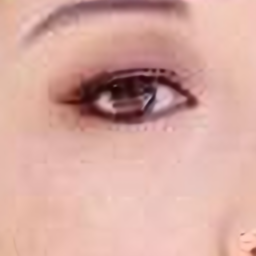}
\caption{EDSR}\label{fig:mouse}
\end{subfigure}\\
\begin{subfigure}[b]{.27\linewidth}
\includegraphics[width=\linewidth]{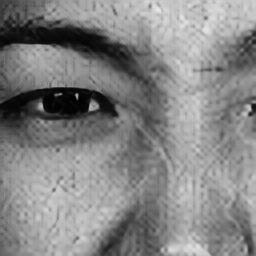}
\caption{SRGAN}\label{fig:mouse}
\end{subfigure}
\begin{subfigure}[b]{.27\linewidth}
\includegraphics[width=\linewidth]{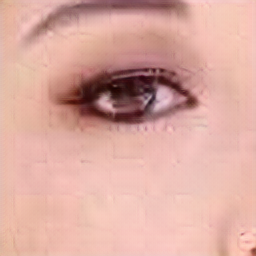}
\caption{SRGAN}\label{fig:mouse}
\end{subfigure}\\
\begin{subfigure}[b]{.27\linewidth}
\includegraphics[width=\linewidth]{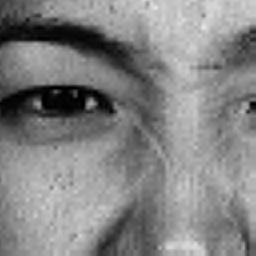}
\caption{Proposed(Daub)}\label{fig:mouse}
\end{subfigure}
\begin{subfigure}[b]{.27\linewidth}
\includegraphics[width=\linewidth]{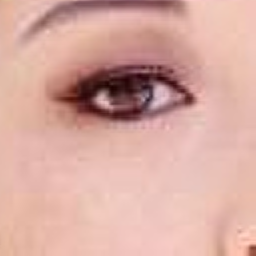}
\caption{Proposed(Daub)}\label{fig:mouse}
\end{subfigure}\\
\begin{subfigure}[b]{.27\linewidth}
\includegraphics[width=\linewidth]{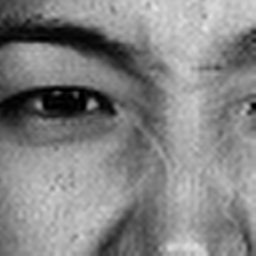}
\caption{Proposed(Meyer)}\label{fig:mouse}
\end{subfigure}
\begin{subfigure}[b]{.27\linewidth}
\includegraphics[width=\linewidth]{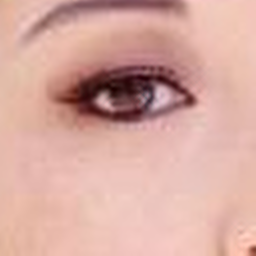}
\caption{Proposed(Meyer)}\label{fig:mouse}
\end{subfigure}

\caption{ Cropped version of  a gray scale face image (left column) and a color face image (right column)  for scaling factor of 4}
\label{Sub_Face4}
\end{figure}

 To quantitatively evaluate the  performance, we calculate the PSNR, SSIM and FSIM values for five types of inputs generated through the five approaches  mentioned in Section \ref{quant}.  We compare  results of our approach with the two best performing existing algorithms found in our earlier  quantitative and subjective evaluations, which are EDSR and SRGAN. We present  the average  PSNR, SSIM and FSIM values over the five types of inputs  for  scaling factors of 2  and  4 in Tables \ref{Jaffe2}-\ref{Face4}. The results show that our proposed approach  performs the best in terms of PSNR and is as good as or better  than the two best performing algorithms EDSR and SRGAN in terms of SSIM and FSIM. For subjective evaluation, we have shown the performance on  cropped versions of a gray scale image taken from  the Jaffe database and a  color image taken from the SET5 database. Figures \ref{Sub_Face22} and  \ref{Sub_Face4} show  the subjective evaluation of the two images for  scaling factors of 2 and 4, respectively. It shows that our proposed approach generates outputs as good as or better than the other two. 

\begin{figure}
\centering
\begin{subfigure}[b]{.14\linewidth}
\includegraphics[width=\linewidth]{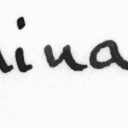}
\caption{Original}\label{fig:mouse}
\end{subfigure}
\hspace{52pt}
\begin{subfigure}[b]{.14\linewidth}
\includegraphics[width=\linewidth]{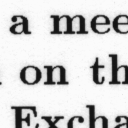}
\caption{Original}\label{fig:mouse}
\end{subfigure}\\
\begin{subfigure}[b]{.27\linewidth}
\includegraphics[width=\linewidth]{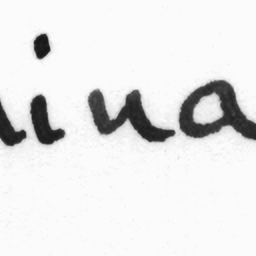}
\caption{EDSR}\label{fig:mouse}
\end{subfigure}
\begin{subfigure}[b]{.27\linewidth}
\includegraphics[width=\linewidth]{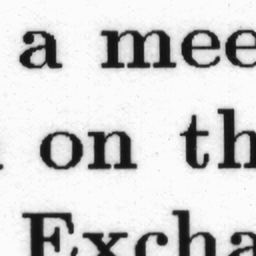}
\caption{EDSR}\label{fig:mouse}
\end{subfigure}\\
\begin{subfigure}[b]{.27\linewidth}
\includegraphics[width=\linewidth]{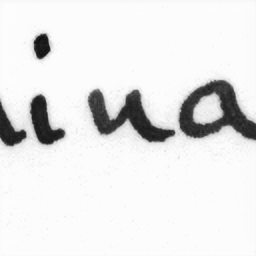}
\caption{SRGAN}\label{fig:mouse}
\end{subfigure}
\begin{subfigure}[b]{.27\linewidth}
\includegraphics[width=\linewidth]{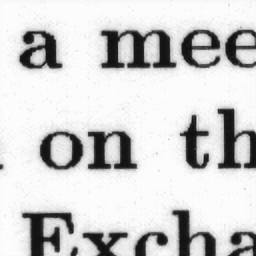}
\caption{SRGAN}\label{fig:mouse}
\end{subfigure}\\
\begin{subfigure}[b]{.27\linewidth}
\includegraphics[width=\linewidth]{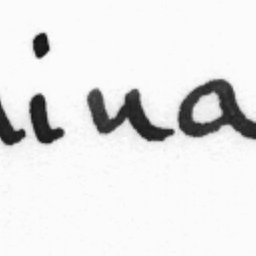}
\caption{Proposed(Daub)}\label{fig:mouse}
\end{subfigure}
\begin{subfigure}[b]{.27\linewidth}
\includegraphics[width=\linewidth]{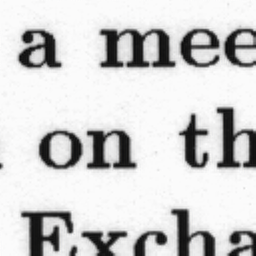}
\caption{Proposed(Daub)}\label{fig:mouse}
\end{subfigure}\\
\begin{subfigure}[b]{.27\linewidth}
\includegraphics[width=\linewidth]{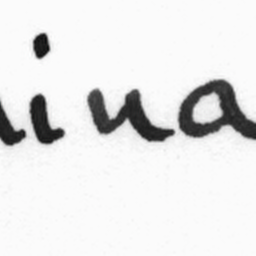}
\caption{Proposed(Meyer)}\label{fig:mouse}
\end{subfigure}
\begin{subfigure}[b]{.27\linewidth}
\includegraphics[width=\linewidth]{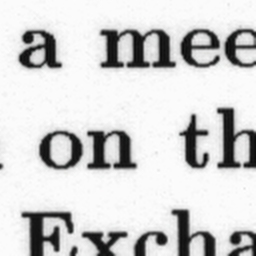}
\caption{Proposed(Meyer)}\label{fig:mouse}
\end{subfigure}

\caption{ Cropped version of  a hand written text image  (left column) and  a machine printed text image (right column)  for scaling factor of 2}
\label{Sub_Text2}
\end{figure}

\begin{figure}
\centering
\begin{subfigure}[b]{.07\linewidth}
\includegraphics[width=\linewidth]{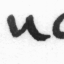}
\caption{Original}\label{fig:mouse}
\end{subfigure}
\hspace{52pt}
\begin{subfigure}[b]{.07\linewidth}
\includegraphics[width=\linewidth]{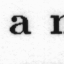}
\caption{Original}\label{fig:mouse}
\end{subfigure}\\
\begin{subfigure}[b]{.27\linewidth}
\includegraphics[width=\linewidth]{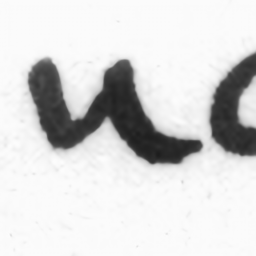}
\caption{EDSR}\label{fig:mouse}
\end{subfigure}
\begin{subfigure}[b]{.27\linewidth}
\includegraphics[width=\linewidth]{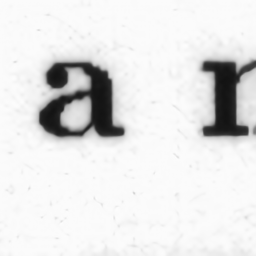}
\caption{EDSR}\label{fig:mouse}
\end{subfigure}\\
\begin{subfigure}[b]{.27\linewidth}
\includegraphics[width=\linewidth]{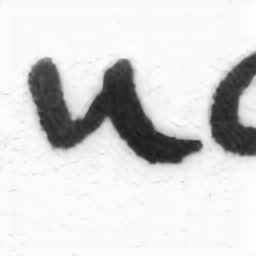}
\caption{SRGAN}\label{fig:mouse}
\end{subfigure}
\begin{subfigure}[b]{.27\linewidth}
\includegraphics[width=\linewidth]{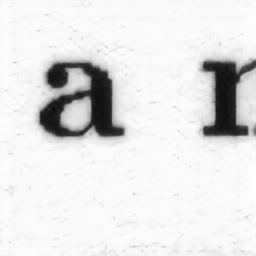}
\caption{SRGAN}\label{fig:mouse}
\end{subfigure}\\
\begin{subfigure}[b]{.27\linewidth}
\includegraphics[width=\linewidth]{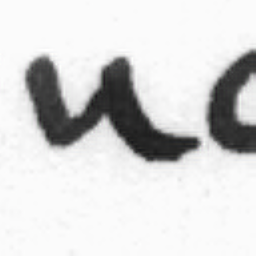}
\caption{Proposed(Daub)}\label{fig:mouse}
\end{subfigure}
\begin{subfigure}[b]{.27\linewidth}
\includegraphics[width=\linewidth]{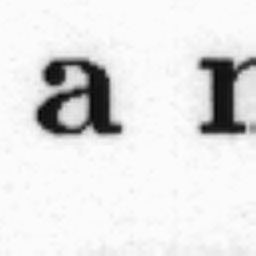}
\caption{Proposed(Daub)}\label{fig:mouse}
\end{subfigure}\\
\begin{subfigure}[b]{.27\linewidth}
\includegraphics[width=\linewidth]{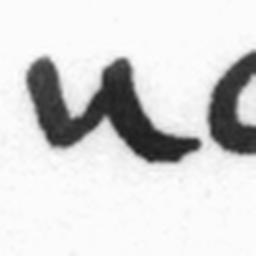}
\caption{Proposed(Meyer)}\label{fig:mouse}
\end{subfigure}
\begin{subfigure}[b]{.27\linewidth}
\includegraphics[width=\linewidth]{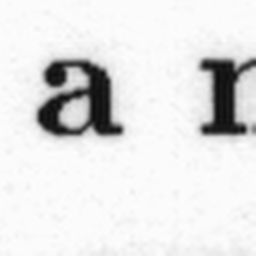}
\caption{Proposed(Meyer)}\label{fig:mouse}
\end{subfigure}

\caption{ Cropped version of  a hand written text image  (left column) and a machine printed text image (right column)  for scaling factor of 4}
\label{Sub_Text4}
\end{figure}

On the other hand, text images are images which mainly have very low and very high frequencies alone unlike most natural images. As pointed out in \cite{walha2016resolution}, text images have sudden discontinuity, regularity in fine pattern and continuity of same intensity values along a particular direction. As our proposed approach is not targeted to work specifically on such special images, we do not expect it to perform well in text images.      For an analysis, we have taken  cropped versions of handwritten and machine printed images from standard IAM\footnote{\url{http://www.fki.inf.unibe.ch/databases/iam-handwriting-database/download-the-iam-handwriting-database}} database to perform subjective comparison.  Figures \ref{Sub_Text2} and  \ref{Sub_Text4}  show the interpolated output for  scaling factors of 2 and 4, respectively. Interestingly, we find qualitatively that the performance of our algorithm is comparable to that of the two best performing existing algorithms EDSR and SRGAN even in the case of text images.
\begin{table}
\caption{PSNR(dB) comparison of gray scale images from USC SIPI Miscellaneous database	for scaling factor 4	}
\begin{center}
\label{Tab_PSNR_GRAY4}
\scalebox{1}{%
\begin{tabular}{|l|l|l|l|l|l|l|l|}
\hline
\multirow{2}{*}{\begin{tabular}[c]{@{}l@{}}Input Low\\  Resolution\end{tabular}} & \multicolumn{7}{c|}{Technique}                                                                                                                   \\ \cline{2-8} 
                                                                                 & NEDI  & SR-TSE & FPCS-LAL & EDSR & SRGAN & \begin{tabular}[c]{@{}l@{}}Proposed\\ (Daub)\end{tabular} & \begin{tabular}[c]{@{}l@{}}Proposed\\ (Meyer)\end{tabular} \\ \hline
Bicubic                                                                          & 26.11 & 27.11  & 26.54    & \textbf{27.92}	& 27.22
 & 25.92                                                   & 26.08                                                      \\ \hline
Daubechies                                                                       & 26.39 & 26.24  & 26.19   & 26.40	& 26.09
  & \textbf{26.47}                                                   & 26.29                                                      \\ \hline
D-Meyer                                                                          & 26.30 & 25.76  & 26.03   & 25.79	& 25.54
  & 26.15                                                   & \textbf{26.70}                                                      \\ \hline
Gaussian                                                                         & 25.92 &  26.29      & 25.92   & \textbf{26.54}	& 26.26
  & 25.89                                                   & 26.13                                                      \\ \hline
Sub-sampling                                                                    & \textbf{25.28} & 21.85  & 23.05    & 21.40	& 21.76
  & 23.69                                                   & 23.81                                                      \\ \hline
Average                                                                          &  \textbf{26.00}     &    25.45    &     25.54    & 25.61	& 25.37
  &                                                        25.62 &   {25.80}                                                          \\ \hline

\end{tabular}}

\end{center}
\end{table}

\begin{table}
\caption{SSIM comparison of gray scale images from USC SIPI  Miscellaneous database	for scaling factor 4		}
\begin{center}
\label{Tab_FSIM_gray4}
\scalebox{1}{%
\begin{tabular}{|l|l|l|l|l|l|l|l|}
\hline
\multirow{2}{*}{\begin{tabular}[c]{@{}l@{}}Input Low\\  Resolution\end{tabular}} & \multicolumn{7}{c|}{Technique}                                                                                                                   \\ \cline{2-8} 
                                                                                 & NEDI  & SR-TSE & FPCS-LAL & EDSR & SRGAN & \begin{tabular}[c]{@{}l@{}}Proposed\\ (Daub)\end{tabular} & \begin{tabular}[c]{@{}l@{}}Proposed\\ (Meyer)\end{tabular} \\ \hline
Bicubic                                                                          & 0.6532 & 0.7291 & 0.7082 & \textbf{0.7565}	& 0.7352
   & 0.6880                                                  & 0.6927                                                     \\ \hline
Daubechies                                                                       & 0.6677 & 0.7213 & 0.7084  & \textbf{0.7439}	& 0.7199
  & 0.7201                                                  & 0.6956                                                     \\ \hline
D-Meyer                                                                          & 0.6633 & 0.7030 & 0.6967 & \textbf{0.7158}	& 0.6929
   & 0.6860                                                  & 0.7103                                                     \\ \hline
Gaussian                                                                         & 0.6414 &  0.7101     & 0.6867  & \textbf{ 0.7231}	& 0.711
  & 0.6709                                                  & 0.6858                                                     \\ \hline
Sub-sampling                                                                    & \textbf{0.6624} & 0.5810 & 0.6194   & 0.5872	& 0.5749
 & 0.6485                                                  & 0.6320                                                     \\ \hline
Average                                                                          &    0.6596    &  {0.6889}      &   0.6839    & \textbf{0.7053}	& 0.6868
    &0.6827                                                         &          0.6833                                                  \\ \hline
 
\end{tabular}}

\end{center}
\end{table}

\begin{table}
\caption{FSIM comparison of gray scale  images from USC SIPI Miscellaneous database	for scaling factor 4		}
\begin{center}
\label{Tab_FSIM_gray4l}
\scalebox{1}{%
\begin{tabular}{|l|l|l|l|l|l|l|l|}
\hline
\multirow{2}{*}{\begin{tabular}[c]{@{}l@{}}Input Low\\  Resolution\end{tabular}} & \multicolumn{7}{c|}{Technique}                                                                                                                   \\ \cline{2-8} 
                                                                                 & NEDI  & SR-TSE & FPCS-LAL & EDSR & SRGAN & \begin{tabular}[c]{@{}l@{}}Proposed\\ (Daub)\end{tabular} & \begin{tabular}[c]{@{}l@{}}Proposed\\ (Meyer)\end{tabular} \\ \hline
Bicubic                                                                          & 0.8553 & 0.9186 & 0.8955  & \textbf{0.9303}	& 0.9257
  & 0.8791                                                  & 0.8783                                                     \\ \hline
Daubechies                                                                       & 0.8686 & 0.9019 & 0.8928   & \textbf{0.9110}	& 0.9078
 & 0.8985                                                  & 0.8848                                                     \\ \hline
D-Meyer                                                                          & 0.8695 & 0.8998 & 0.8937  & \textbf{0.9044}	& 0.8961
  & 0.8940                                                  & 0.8938                                                     \\ \hline
Gaussian                                                                         & 0.8418 & 0.9048 & 0.8847  & \textbf{0.9125}	& 0.9056
  & 0.8768                                                  & 0.8750                                                     \\ \hline
Sub-sampling                                                                    & \textbf{0.8673} & 0.8208 & 0.8383  & 0.8214	& 0.8223
  & 0.8544                                                  & 0.8420                                                     \\ \hline
Average                                                                          & 0.8605       &     0.8892   &    0.8810   & \textbf{0.8959}	& {0.8915}
    &                                                        0.8801 &    0.8748                                                        \\ \hline

\end{tabular}}

\end{center}
\end{table}

\begin{table}
\caption{PSNR(dB) comparison of color scale images from USC SIPI Miscellaneous database	for scaling factor 4		}
\begin{center}
\label{Tab_PSNR_USCCOLOR4}
\scalebox{1}{%
\begin{tabular}{|l|l|l|l|l|l|l|l|}
\hline
\multirow{2}{*}{\begin{tabular}[c]{@{}l@{}}Input Low\\  Resolution\end{tabular}} & \multicolumn{7}{c|}{Technique}                                                                                                                   \\ \cline{2-8} 
                                                                                 & NEDI  & SR-TSE & FPCS-LAL & EDSR & SRGAN & \begin{tabular}[c]{@{}l@{}}Proposed\\ (Daub)\end{tabular} & \begin{tabular}[c]{@{}l@{}}Proposed\\ (Meyer)\end{tabular} \\ \hline
Bicubic                                                                          & 27.9  & 30.96  & 30.00      & \textbf{32.37}	& 30.99
  & 28.78                                                   & 29                                                         \\ \hline
Daubechies                                                                       & 28.08 & 29.40  & 29.19    & 29.36	& 28.91
 & 29.46                                                   & \textbf{29.58}                                                      \\ \hline
D-Meyer                                                                          & 28.03 & 28.67  & 28.90   & 28.33	& 28.00
  & 29.26                                                   & \textbf{30.04}                                                      \\ \hline
Gaussian                                                                         & 27.61 & 29.43  & 28.82   & \textbf{29.94}	& 29.31
  & 29.00                                                      & 29.14                                                      \\ \hline
Sub-sampling                                                                    & 27.07 & 25.21  & 26.3    & 24.29	& 24.54
  & 27.31                                                   & \textbf{27.42}                                                      \\ \hline
Average                                                                          &   27.74    & 28.73       &    28.64      &                                                   28.86	& 28.35
 &  {28.76} &  \textbf{29.04}                                                          \\ \hline

\end{tabular}}

\end{center}
\end{table}

\begin{table}
\caption{PSNR(dB) comparison of color scale images from BSD  database for scaling factor 4			}
\begin{center}
\label{Tab_PSNR_BSD4}
\scalebox{1}{%
\begin{tabular}{|l|l|l|l|l|l|l|l|}
\hline
\multirow{2}{*}{\begin{tabular}[c]{@{}l@{}}Input Low\\  Resolution\end{tabular}} & \multicolumn{7}{c|}{Technique}                                                                                                                   \\ \cline{2-8} 
                                                                                 & NEDI  & SR-TSE & FPCS-LAL & EDSR & SRGAN & \begin{tabular}[c]{@{}l@{}}Proposed\\ (Daub)\end{tabular} & \begin{tabular}[c]{@{}l@{}}Proposed\\ (Meyer)\end{tabular} \\ \hline
Bicubic                                                                          & 25.34 & 26.85  & 26.39  & \textbf{ 27.58}	& 27.02
  & 25.84                                                   & 25.91                                                      \\ \hline
Daubechies                                                                       & 25.58 & 26.02  & 26.03  & 26.06	& 25.92
  & 26.18                                                   & \textbf{26.21}                                                      \\ \hline
D-Meyer                                                                          & 25.57 & 25.69  & 25.90  & 25.51	& 25.42
  & 26.06                                                   & \textbf{26.49}                                                      \\ \hline
Gaussian                                                                         & 24.69 & 26.09  & 25.82   & \textbf{26.44}	& 26.12
 & 25.90                                                   & 25.95                                                      \\ \hline
Sub-sampling                                                                    & \textbf{24.51} & 21.69  & 22.91   & 20.83	& 21.42
 & 23.73                                                   & 23.69                                                      \\ \hline
Average                                                                          & 25.14      &  25.27      &    25.41     & 25.28	& 25.18
 &                                                        {25.54} &  \textbf{25.65}                                                          \\ \hline

\end{tabular}}

\end{center}
\end{table}

\begin{table}
\caption{PSNR(dB) comparison of color images from SET 5 database for scaling factor 4	}
\begin{center}
\label{Tab_PSNR_Set5_4}
\scalebox{0.9}{%

\begin{tabular}{|l|l|l|l|l|l|l|l|}
\hline
\multirow{2}{*}{\begin{tabular}[c]{@{}l@{}}Input Low\\  Resolution\end{tabular}} & \multicolumn{7}{c|}{Technique}                                                                                                                   \\ \cline{2-8} 
                                                                                 & NEDI  & SR-TSE & FPCS-LAL & EDSR & SRGAN & \begin{tabular}[c]{@{}l@{}}Proposed\\ (Daub)\end{tabular} & \begin{tabular}[c]{@{}l@{}}Proposed\\ (Meyer)\end{tabular}  \\ \hline
Bicubic      & 27.09 & 30.21  & 29.26    & \textbf{31.98} & 30.57 & 27.96 & 28.21\\ \hline
Daubechies   & 27.40  & 28.6   & 28.46    & 28.7  & 28.32 & 28.79 & \textbf{29.06} \\ \hline
D-Meyer      & 27.26 & 27.77  & 28.12    & 27.51 & 27.24 & 28.66 & \textbf{29.48} \\ \hline
Gaussian     & 26.73 & 28.51  & 28.00       & \textbf{29.16} & 28.58 & 28.17 & 28.31 \\ \hline
Sub-sampling & 26.64 & 24.34  & 25.62    & 23.12 & 23.63 & 26.76 & \textbf{26.94}\\ \hline
Average      & 27.02 & 27.89  & 27.89    & {28.10}  & 27.67 & 28.07 & \textbf{28.40} \\ \hline
 
\end{tabular}}

\end{center}
\end{table}

\begin{table}
\caption{PSNR(dB) comparison of gray scale and color images from SET 14 database for  scaling factor 4	}
\begin{center}
\label{Tab_PSNR_Set14_4}
\scalebox{0.9}{%

\begin{tabular}{|l|l|l|l|l|l|l|l|}
\hline
\multirow{2}{*}{\begin{tabular}[c]{@{}l@{}}Input Low\\  Resolution\end{tabular}} & \multicolumn{7}{c|}{Technique}                                                                                                                   \\ \cline{2-8} 
                                                                                 & NEDI  & SR-TSE & FPCS-LAL & EDSR & SRGAN & \begin{tabular}[c]{@{}l@{}}Proposed\\ (Daub)\end{tabular} & \begin{tabular}[c]{@{}l@{}}Proposed\\ (Meyer)\end{tabular}  \\ \hline
Bicubic      & 23.47 & 27.39  & 26.51    & \textbf{28.33} & 27.49 & 25.65 & 25.83 \\ \hline
Daubechies   & 25.34 & 26.16  & 26.05    & 25.90 & 25.82 & 26.26 & \textbf{26.30} \\ \hline
D-Meyer      & 25.20 & 25.47  & 25.81    & 25.19 & 24.99 & 26.16 & \textbf{26.62} \\ \hline
Gaussian     & 24.82 & 26.27  & 25.70    & \textbf{26.58} & 26.15 & 25.83 & 25.88 \\ \hline
Sub-sampling & \textbf{24.21} & 19.93  & 23.03    & 20.54 & 21.12 & 24.01 & 23.97 \\ \hline
Average      & 24.61 & 25.04  & 25.42    & 25.31 & 25.11 & {25.58} & \textbf{25.72}  \\ \hline
 
\end{tabular}}

\end{center}
\end{table}

\begin{table}
\caption{SSIM comparison of color  images from USC SIPI  Miscellaneous database	for scaling factor 4		}
\begin{center}
\label{Tab_SSIM_USCCOLOR4}
\scalebox{1}{%
\begin{tabular}{|l|l|l|l|l|l|l|l|}
\hline
\multirow{2}{*}{\begin{tabular}[c]{@{}l@{}}Input Low\\  Resolution\end{tabular}} & \multicolumn{7}{c|}{Technique}                                                                                                                   \\ \cline{2-8} 
                                                                                 & NEDI  & SR-TSE & FPCS-LAL & EDSR & SRGAN & \begin{tabular}[c]{@{}l@{}}Proposed\\ (Daub)\end{tabular} & \begin{tabular}[c]{@{}l@{}}Proposed\\ (Meyer)\end{tabular} \\ \hline
Bicubic                                                                          & 0.7895 & 0.8493 & 0.8289   & \textbf{0.8697}	& 0.8399
    & 0.8104                                                  & 0.8127                                                     \\ \hline
Daubechies                                                                       & 0.8003 & 0.8358 & 0.8229   &\textbf{0.8497}	& 0.8174
    & 0.8232                                                  & 0.8180                                                     \\ \hline
D-Meyer                                                                          & 0.7950 & 0.8162 & 0.8098   & \textbf{0.8207}	& 0.7823
    & 0.8064                                                  & 0.8201                                                     \\ \hline
Gaussian                                                                         & 0.7825 & 0.8287 & 0.8118   & \textbf{ 0.8416}	& 0.8199
    & 0.8096                                                  & 0.8123                                                     \\ \hline
Sub-sampling                                                                    & \textbf{0.7964} & 0.7392 & 0.7598   & 0.7323	& 0.6937
    & 0.7857                                                  & 0.7718                                                     \\ \hline
Average                                                                          &     0.7927   &    {0.8138}    &   0.8067      & \textbf{0.8228}	& 0.7906
     &                                                        0.8070 &   0.8070                                                         \\ \hline

\end{tabular}}

\end{center}
\end{table}

\begin{table}
\caption{SSIM comparison of color  images from BSD  database	for scaling factor 4		}
\begin{center}
\label{Tab_SSIM_BSD4}
\scalebox{1}{%
\begin{tabular}{|l|l|l|l|l|l|l|l|}
\hline
\multirow{2}{*}{\begin{tabular}[c]{@{}l@{}}Input Low\\  Resolution\end{tabular}} & \multicolumn{7}{c|}{Technique}                                                                                                                   \\ \cline{2-8} 
                                                                                 & NEDI  & SR-TSE & FPCS-LAL & EDSR & SRGAN & \begin{tabular}[c]{@{}l@{}}Proposed\\ (Daub)\end{tabular} & \begin{tabular}[c]{@{}l@{}}Proposed\\ (Meyer)\end{tabular} \\ \hline
Bicubic                                                                          & 0.6400 & 0.7121 & 0.6923   & \textbf{0.7362}	& 0.7124
 & 0.6738                                                  & 0.6735                                                     \\ \hline
Daubechies                                                                       & 0.6531 & 0.7008 & 0.69132  & \textbf{0.7204}	& 0.6965
 & 0.6914                                                  & 0.6853                                                     \\ \hline
D-Meyer                                                                          & 0.6495 & 0.6878 & 0.6827   & \textbf{ 0.6959}	& 0.6684
 & 0.6782                                                  & 0.6941                                                     \\ \hline
Gaussian                                                                         & 0.6145 & 0.6838 & 0.6686   & \textbf{0.7000}	& 0.6829
 & 0.6677                                                  & 0.6677                                                     \\ \hline
Sub-sampling                                                                    & \textbf{0.6495} & 0.5816 & 0.6105   & 0.5735	& 0.5615
 & 0.6444                                                  & 0.6266                                                     \\ \hline
Average                                                                          &   0.6413     & {0.6732}       & 0.6690         & \textbf{0.6852}	& 0.6643
 & 0.6711 &      0.6694                                                      \\ \hline

\end{tabular}}

\end{center}
\end{table}

\begin{table}
\caption{SSIM comparison of color images from SET 5 database for scaling factor 4}
\begin{center}
\label{Tab_SSIM_Set5_4}
\scalebox{0.9}{%

\begin{tabular}{|l|l|l|l|l|l|l|l|}
\hline
\multirow{2}{*}{\begin{tabular}[c]{@{}l@{}}Input Low\\  Resolution\end{tabular}} & \multicolumn{7}{c|}{Technique}                                                                                                                   \\ \cline{2-8} 
                                                                                 & NEDI  & SR-TSE & FPCS-LAL & EDSR & SRGAN & \begin{tabular}[c]{@{}l@{}}Proposed\\ (Daub)\end{tabular} & \begin{tabular}[c]{@{}l@{}}Proposed\\ (Meyer)\end{tabular}  \\ \hline
Bicubic      & 0.7792 & 0.8593 & 0.8272   & 0.8048 & \textbf{0.8598} & 0.8029 & 0.8088 \\ \hline
Daubechies   & 0.7917 & 0.8388 & 0.8167   & \textbf{0.8566} & 0.8229 & 0.8173 & 0.8195 \\ \hline
D-Meyer      & 0.7823 & 0.8147 & 0.8012   & 0.8173 & 0.7790 & 0.8029 & \textbf{0.8220} \\ \hline
Gaussian     & 0.7658 & 0.8297 & 0.8074   & \textbf{0.8524} & 0.8291 & 0.8033 & 0.8075 \\ \hline
Sub-sampling & \textbf{0.7910} & 0.7260 & 0.747    & 0.7044 & 0.6803 & 0.7848 & 0.7694 \\ \hline
Average      & 0.7820 & \textbf{0.8137} & 0.7999   & {0.8071} & 0.7942 & 0.8022 & 0.8054 \\ \hline
 
\end{tabular}}

\end{center}
\end{table}

\begin{table}
\caption{SSIM comparison of gray scale and color images from SET 14 database for scaling factor 4	}
\begin{center}
\label{Tab_SSIM_Set14_4}
\scalebox{0.9}{%

\begin{tabular}{|l|l|l|l|l|l|l|l|}
\hline
\multirow{2}{*}{\begin{tabular}[c]{@{}l@{}}Input Low\\  Resolution\end{tabular}} & \multicolumn{7}{c|}{Technique}                                                                                                                   \\ \cline{2-8} 
                                                                                 & NEDI  & SR-TSE & FPCS-LAL & EDSR & SRGAN & \begin{tabular}[c]{@{}l@{}}Proposed\\ (Daub)\end{tabular} & \begin{tabular}[c]{@{}l@{}}Proposed\\ (Meyer)\end{tabular}  \\ \hline
Bicubic      & 0.6260  & 0.7518 & 0.7243   & \textbf{0.7785} & 0.7496 & 0.7009 & 0.7025 \\ \hline
Daubechies   & 0.6878 & 0.7353 & 0.7207   & \textbf{0.7506} & 0.7231 & 0.7218 & 0.7135 \\ \hline
D-Meyer      & 0.6826 & 0.7159 & 0.7083   & \textbf{0.7208} & 0.6866 & 0.7072 & 0.7205 \\ \hline
Gaussian     & 0.6609 & 0.7169 & 0.6997   & \textbf{0.7379} & 0.7176 & 0.6979 & 0.6970  \\ \hline
Sub-sampling & \textbf{0.6845} & 0.5321 & 0.6398   & 0.5928 & 0.5743 & 0.6776 & 0.6539 \\ \hline
Average      & 0.6684 & 0.6904 & 0.6986   & \textbf{0.7161} & 0.6902 & {0.7011} & 0.6975 \\ \hline
 
\end{tabular}}

\end{center}
\end{table}

\begin{table}
\caption{FSIM comparison of color  images from USC SIPI  Miscellaneous database	for scaling factor 4		}
\begin{center}
\label{Tab_FSIM_USCCOLOR4}
\scalebox{1}{%
\begin{tabular}{|l|l|l|l|l|l|l|l|}
\hline
\multirow{2}{*}{\begin{tabular}[c]{@{}l@{}}Input Low\\  Resolution\end{tabular}} & \multicolumn{7}{c|}{Technique}                                                                                                                   \\ \cline{2-8} 
                                                                                 & NEDI  & SR-TSE & FPCS-LAL & EDSR & SRGAN & \begin{tabular}[c]{@{}l@{}}Proposed\\ (Daub)\end{tabular} & \begin{tabular}[c]{@{}l@{}}Proposed\\ (Meyer)\end{tabular} \\ \hline
Bicubic                                                                          & 0.8751 & 0.9301 & 0.9092   & \textbf{0.9465}	& 0.9364
 & 0.8948                                                  & 0.8961                                                     \\ \hline
Daubechies                                                                       & 0.8874 & 0.9171 & 0.9051  & \textbf{0.9257}	& 0.9181
  & 0.9054                                                  & 0.9027                                                     \\ \hline
D-Meyer                                                                          & 0.8865 & 0.8995 & 0.8958   & 0.9039	& 0.8924
 & 0.8951                                                  & \textbf{0.9067}                                                     \\ \hline
Gaussian                                                                         & 0.8701 & 0.9117 & 0.8971  & \textbf{0.9215} &	0.9161
  & 0.8948                                                  & 0.8968                                                     \\ \hline
Sub-sampling                                                                    & \textbf{0.8888} & 0.8500 & 0.8639   & 0.8466	& 0.8358
 & 0.8816                                                  & 0.8754                                                     \\ \hline
Average                                                                          &    0.8816     &  {0.9017}      &  0.8942   & \textbf{0.9088}	& 0.8998
      &                                                        0.8943 &      0.8955                                                      \\ \hline

\end{tabular}}

\end{center}
\end{table}

\begin{table} 
\caption{FSIM comparison of color  images from BSD database		for scaling factor 4	}
\begin{center}
\label{Tab_FSIM_BSD4}
\scalebox{1}{%
\begin{tabular}{|l|l|l|l|l|l|l|l|}
\hline
\multirow{2}{*}{\begin{tabular}[c]{@{}l@{}}Input Low\\  Resolution\end{tabular}} & \multicolumn{7}{c|}{Technique}                                                                                                                   \\ \cline{2-8} 
                                                                                 & NEDI  & SR-TSE & FPCS-LAL & EDSR & SRGAN & \begin{tabular}[c]{@{}l@{}}Proposed\\ (Daub)\end{tabular} & \begin{tabular}[c]{@{}l@{}}Proposed\\ (Meyer)\end{tabular} \\ \hline
Bicubic                                                                          & 0.7435 & 0.8246 & 0.8038   & \textbf{0.8378}	& 0.8361
 & 0.7874                                                  & 0.7870                                                     \\ \hline
Daubechies                                                                       & 0.7611 & 0.8286 & 0.8165  & \textbf{0.836}	& 0.8351
  & 0.8063                                                  & 0.8052                                                     \\ \hline
D-Meyer                                                                          & 0.7643 & 0.8204 & 0.8128  & \textbf{0.8243}	& 0.8205
  & 0.8024                                                  & 0.8107                                                     \\ \hline
Gaussian                                                                         & 0.7200 & 0.7985 & 0.7795 & 0.8048	& \textbf{0.8088}
   & 0.7733                                                  & 0.7721                                                     \\ \hline
Sub-sampling                                                                    & 0.7932 & 0.7764 & 0.8031  & 0.7588	& 0.7677
 & \textbf{0.8158}                                                  & 0.8080                                                     \\ \hline
Average                                                                          &  0.7564     &     0.8093   &     0.8031     &                                                      {0.8123}	& \textbf{0.8137}
 &  0.7970 &    0.7966                                                        \\ \hline

\end{tabular}}

\end{center}
\end{table}

\begin{table}
\caption{FSIM comparison of color images from SET 5 database for scaling factor 4	}
\begin{center}
\label{Tab_FSIM_Set5_4}
\scalebox{0.9}{%

\begin{tabular}{|l|l|l|l|l|l|l|l|}
\hline
\multirow{2}{*}{\begin{tabular}[c]{@{}l@{}}Input Low\\  Resolution\end{tabular}} & \multicolumn{7}{c|}{Technique}                                                                                                                   \\ \cline{2-8} 
                                                                                 & NEDI  & SR-TSE & FPCS-LAL & EDSR & SRGAN & \begin{tabular}[c]{@{}l@{}}Proposed\\ (Daub)\end{tabular} & \begin{tabular}[c]{@{}l@{}}Proposed\\ (Meyer)\end{tabular}  \\ \hline
Bicubic      & 0.838  & 0.9004 & 0.7963   & 0.8682 & \textbf{0.9203} & 0.8582 & 0.8607 \\ \hline
Daubechies   & 0.8502 & 0.8870 & 0.8700   & \textbf{0.9012} & 0.8897 & 0.8685 & 0.8700\\ \hline
D-Meyer      & 0.8455 & 0.8713 & 0.8605   & \textbf{0.8798} & 0.8658 & 0.8599 & 0.8722 \\ \hline
Gaussian     & 0.8317 & 0.8817 & 0.8613   & \textbf{0.9038} & 0.8967 & 0.8576 & 0.8595 \\ \hline
Sub-sampling & \textbf{0.8660}& 0.8236 & 0.8406   & 0.8067 & 0.8049 & 0.8641 & 0.8538 \\ \hline
Average      & 0.8463 & {0.8728} & 0.8457   & 0.8719 & \textbf{0.8755} & 0.8617 & 0.8632 \\ \hline
 
\end{tabular}}

\end{center}
\end{table}

\begin{table}
\caption{FSIM comparison of gray scale and color images from SET 14 database for scaling factor 4	}
\begin{center}
\label{Tab_FSIM_Set14_4}
\scalebox{0.9}{%

\begin{tabular}{|l|l|l|l|l|l|l|l|}
\hline
\multirow{2}{*}{\begin{tabular}[c]{@{}l@{}}Input Low\\  Resolution\end{tabular}} & \multicolumn{7}{c|}{Technique}                                                                                                                   \\ \cline{2-8} 
                                                                                 & NEDI  & SR-TSE & FPCS-LAL & EDSR & SRGAN & \begin{tabular}[c]{@{}l@{}}Proposed\\ (Daub)\end{tabular} & \begin{tabular}[c]{@{}l@{}}Proposed\\ (Meyer)\end{tabular}  \\ \hline
Bicubic      & 0.8010 & 0.9005 & 0.8745   & \textbf{0.9172} & 0.9097 & 0.8622 & 0.8605 \\ \hline
Daubechies   & 0.8465 & 0.8874 & 0.8743   & \textbf{0.8917} & 0.8879 & 0.8753 & 0.8716 \\ \hline
D-Meyer      & 0.8482 & 0.8712 & 0.8665   & 0.8725 & 0.8646 & 0.8706 & \textbf{0.8771} \\ \hline
Gaussian     & 0.8190 & 0.8595 & 0.8554   & \textbf{0.8833} & 0.8809 & 0.8546 & 0.8530 \\ \hline
Sub-sampling & \textbf{0.8539} & 0.7665 & 0.8291   & 0.7907 & 0.7915 & \textbf{0.8539} & 0.8414 \\ \hline
Average      & 0.8337 & 0.8570 & 0.8600   & \textbf{0.8711} & {0.8669} & 0.8633 & 0.8607 \\ \hline
 
\end{tabular}}

\end{center}
\end{table}

\begin{table}[]
\centering
\caption{Computation Time (CPU time) of Proposed (Daub) obtained by compromising around $10^{-2}$  dB in PSNR for USC SIPI Miscellaneous gray scale image database for input generated by Wavelet filter approximation (Daubechies)}
\label{Time_Aux}
\begin{tabular}{|l|l|l|l|}
\hline
\multirow{2}{*}{\begin{tabular}[c]{@{}l@{}}Scaling\\ Factor\end{tabular}} & \multicolumn{3}{c|}{Input Image Size (Time in Seconds)}       \\ \cline{2-4} 
                                                                          & $64\times 64$   & \multicolumn{1}{c|}{$128\times 128$} & $256 \times 256$ \\ \hline
$ \times 2                                                                  $ & \makecell{0.215
 \\ $\pm$ 0.010}   &        \makecell{0.312
 \\ $\pm$ 0.021}                  & \makecell{0.645
 \\ $\pm$ 0.031}  \\ \hline
$\times 4 $  & \makecell{0.69
 \\ $\pm$ 0.044}   & \makecell{1.12
 \\ $\pm$ 0.039}      & \makecell{3.79
 \\ $\pm$ 0.051} \\ \hline
\end{tabular}
\end{table}

 \begin{table}
\caption{Computational efficiency analysis on USC SIPI Miscellaneous gray image database for scaling factor 2 }
\label{fom}
\scalebox{0.8}{
\begin{tabular}{|l|l|l|l|l|l|l|l|l|}
\hline
\multirow{2}{*}{Computational Efficiency} & \multicolumn{8}{c|}{Technique}                                                                                                                                                                                                                                                                              \\ \cline{2-9} 
                                  & NEDI   & NARM   & SR-TSE & FPCS-LAL & EDSR     & SRGAN    & \begin{tabular}[c]{@{}l@{}}Proposed\\ (Daub)\end{tabular} & \begin{tabular}[c]{@{}l@{}}Proposed\\ (Meyer)\end{tabular} \\ \hline
Avg. Time(sec)                    & 14.36     & 403    & 210    & 526 &     GPU   &    GPU   & 1.29 & 3.76 \\ \hline
Avg. PSNR                          & 28.28  & 29     & 28.02  & 28.06    & 28.12862 & 27.42518 & 29.03                                                   & 29.22                                                      \\ \hline
Avg. SSIM                          & 0.8223 & 0.835  & 0.8208 & 0.8055   & 0.826608 & 0.806168 & 0.8495                                                  & 0.8472                                                     \\ \hline
Avg. FSIM                          & 0.9437 & 0.9499 & 0.9424 & 0.9405   & 0.944217 & 0.941171 & 0.9534                                                  & 0.9487                                                     \\ \hline
\end{tabular}}
\end{table}
\section{Conclusion}
\label{Conc}
The paper proposes an approach that performs interpolation  exploiting across-scale `process similarity' in wavelet decomposition. The first stage of our proposed approach is the modeling stage, where a model capturing the structural relation between scales is estimated. The model is in the form of optimal weights obtained through particle swarm optimization (PSO).  The weights are used to fuse the high frequency components that are subsequently employed to estimate the input image from its low resolution approximation. The second stage is the generation stage, where these optimal weights modeling the structural relation between scales are used to generate the higher resolution image from the input image.  In this stage, the approach considers  the input image as the low resolution approximation of the output high resolution image to be generated.

We find through qualitative analysis that our process similarity based interpolation approach produces results with fewer artifacts. Quantitative analysis shows that on an average, our proposed approach, in terms of PSNR, SSIM and FSIM perform either the best or near-best considering different databases containing color and gray scale images, and different scaling factors. The major advantage of our proposed approach  is that it is very fast  in terms of CPU  computation time, much faster than all the  algorithms compared.  

The proposed approach has been implemented by performing both DWT and SWT based on the findings that they complement each other where the former provides value faithfulness and the latter provides accurate location correspondence. The implementation  of our approach has been performed considering two different wavelet functions  (Daubechies wavelet with support 4  and discrete approximation of Meyer filter with support  102) yielding two proposed algorithms. Subjective and quantitative comparisons between these algorithms have shown one of them to be marginally more preferable.
 
The modeling performed in this paper employing process similarity is based on wavelet decomposition. In future, process similarity can be explored as an independent concept beyond the use of wavelet decomposition.

\begin{table}[]
\caption{Quantitative evaluation of gray scale face images from Jaffe database for scaling factor 2	}
\begin{center}
\begin{tabular}{|c|l|l|l|l|}
\hline
\multirow{2}{*}{Measure}    & \multicolumn{4}{c|}{Technique}                                                                                                                               \\ \cline{2-5} 
                           & EDSR   & SRGAN  & \begin{tabular}[c]{@{}l@{}}Proposed\\ (Daub)\end{tabular} & \multicolumn{1}{c|}{\begin{tabular}[c]{@{}c@{}}Proposed\\ (Meyer)\end{tabular}} \\ \hline
\multicolumn{1}{|l|}{PSNR}                       & 30.37  & 28.93  & 30.75                                                     & \textbf{30.93}                                                                  \\ \hline
\multicolumn{1}{|l|}{SSIM} & 0.8352 & 0.7833 & \textbf{0.8417}                                           & 0.8392                                                                          \\ \hline
\multicolumn{1}{|l|}{FSIM} & 0.9358 & 0.9152 & \textbf{0.9384}                                           & 0.9376                                                                          \\ \hline
\end{tabular}
\end{center}
\label{Jaffe2}
\end{table}

\begin{table}[]
\caption{Quantitative evaluation of color face images from SET5  for scaling factor 2	}
\begin{center}
\begin{tabular}{|c|l|l|l|l|}
\hline
\multirow{2}{*}{Measure}    & \multicolumn{4}{c|}{Technique}                                                                                                                               \\ \cline{2-5} 
                           & EDSR   & SRGAN  & \begin{tabular}[c]{@{}l@{}}Proposed\\ (Daub)\end{tabular} & \multicolumn{1}{c|}{\begin{tabular}[c]{@{}c@{}}Proposed\\ (Meyer)\end{tabular}} \\ \hline
\multicolumn{1}{|l|}{PSNR}                       & 33.00  & 31.59  & 33.64                                                     & \textbf{34.11}                                                                  \\ \hline
\multicolumn{1}{|l|}{SSIM} & \textbf{0.9367} & 0.8688 & 0.9139                                           & 0.9162                                                                          \\ \hline
\multicolumn{1}{|l|}{FSIM} & 0.9579 & 0.9402 & 0.9593                                           & \textbf{0.9602}                                                                          \\ \hline
\end{tabular}
\end{center}
\label{Face2}
\end{table}

\begin{table}[]
\caption{Quantitative evaluation of gray scale face images from Jaffe  for scaling factor 4	}
\begin{center}

\begin{tabular}{|c|l|l|l|l|}
\hline
\multirow{2}{*}{Measure}    & \multicolumn{4}{c|}{Technique}                                                                                                                                        \\ \cline{2-5} 
                           & EDSR            & SRGAN  & \begin{tabular}[c]{@{}l@{}}Proposed\\ (Daub)\end{tabular} & \multicolumn{1}{c|}{\begin{tabular}[c]{@{}c@{}}Proposed\\ (Meyer)\end{tabular}} \\ \hline
\multicolumn{1}{|l|}{PSNR}                       & 27.41           & 27.10  & 27.59                                                     & \textbf{27.85}                                                                  \\ \hline
\multicolumn{1}{|l|}{SSIM} & \textbf{0.7516} & 0.7234 & 0.7508                                                    & 0.7506                                                                          \\ \hline
\multicolumn{1}{|l|}{FSIM} & \textbf{0.8780} & \textbf{0.8780} & 0.8730                                                    & 0.8740                                                                          \\ \hline
\end{tabular}

\end{center}
\label{Jaffe4}
\end{table}

\begin{table}[]
\caption{Quantitative evaluation of color  face images from SET5  for scaling factor 4	}
\begin{center}
\begin{tabular}{|c|l|l|l|l|}
\hline
\multirow{2}{*}{Measure}    & \multicolumn{4}{c|}{Technique}                                                                                                                                        \\ \cline{2-5} 
                           & EDSR            & SRGAN  & \begin{tabular}[c]{@{}l@{}}Proposed\\ (Daub)\end{tabular} & \multicolumn{1}{c|}{\begin{tabular}[c]{@{}c@{}}Proposed\\ (Meyer)\end{tabular}} \\ \hline
\multicolumn{1}{|l|}{PSNR}                       & 28.91           & 28.88  & 29.64                                                     & \textbf{29.94}                                                                  \\ \hline
\multicolumn{1}{|l|}{SSIM} & \textbf{0.8447} & 0.7904 & 0.8101                                                    & 0.8134                                                                          \\ \hline
\multicolumn{1}{|l|}{FSIM} & \textbf{0.8998} & 0.8981 & 0.8960                                                    & 0.8972                                                                          \\ \hline
\end{tabular}
\end{center}
\label{Face4}
\end{table}

\section{Acknowledgment}
Authors would like to thank Mr. Debanjan Sengupta and Ms. Anusha Vupputuri, Ph.D. students at IIT Kharagpur, for their help in completing this article. The authors would also like to thank the anonymous reviewers, as their contribution to bring the article to its best version is enormous.

\section*{References}

\bibliography{Biblio}
\appendix
\section{Discrete and Stationary Wavelet Transforms}
\label{app1}
Wavelet transform \cite{mallat1999wavelet} is a powerful multi-resolution analysis (MRA) \cite{mallat1989theory}, designed to obtain the components of input signal at every location in different scales /resolutions. The discrete wavelet transform (DWT) \cite{mallat1989theory} works on discrete signals to produce discrete coefficients quantifying the said components. A discrete signal can be represented as follows
\begin{equation}
T(x) = \sum\limits_P {{a_P}{\beta _P}(x)} 
\end{equation}
 With unique real valued expansion coefficients $a_P$, a $\beta _P(x)$ acts as a basis function over which the signal is represented. In other words, given all the $a_P$s and the corresponding set of basis functions, the signal $T(x)$ can be generated.

In the field of wavelet-based analysis and synthesis, the basis functions are of two types, known as scaling function and wavelet function. The scaling functions defined as $\varphi _P(x)$s  satisfy certain properties. In fact, $P=(s,k)$, where $s,k \in \mathbb{Z^*}$, respectively represent scale and translation, and $\varphi _{(s,k)}(x)=2^{s/2}\varphi (2^sx-k)$. 

One property satisfied by scaling functions, which is  the heart of discrete wavelet transform, is
\begin{equation}
\varphi (x - k) = \sqrt 2 \sum\limits_n {{c_n}} \varphi (2(x - k) - n)
\end{equation} 
where $k,n \in \mathbb{Z^*}$ and $c_n$s are known as scaling function coefficients. Note that the above expression relates $\varphi _P(x)$s at consecutive scales. Another function called the wavelet function ($\psi_P(x)$) is also defined, which is related to the scaling function as follows
\begin{equation}
\label{approx}
\langle {\varphi _P}(x),{\psi _Q}(x)\rangle  = 0\,
\end{equation}
\begin{equation}
\label{deails}
\psi (x - k) = \sqrt 2 \sum\limits_n {{d_n}} \varphi (2(x - k) - n)
\end{equation}
where $d_n$s are known as wavelet function coefficients. Note that, the above expression relates $\varphi _P(x)$ and  $\psi_Q(x)$ at consecutive scales with $\varphi _P(x)$ being at the finer scale.
 In such condition, any function  $T(x) \in {L^2}(\mathbb{R})$ can be decomposed to approximation coefficients (low frequency component) with scaling function and detail coefficients (high frequency component) with wavelet function. They are collectively referred to as the wavelet coefficients.
 
 Now let us consider a discrete signal $T(x)$ having M samples such that $M=2^j,\ j\in\mathbb{Z^*}$. So, $j$ is the highest (finest) possible scale and $0$ is the lowest (coarsest) possible scale. As the input signal $T(x)$ is at the $j^{\textrm{th}}$ scale, it is decomposed into approximation and detail coefficients at scale $j-1$ as follows

\begin{equation}
{W_\varphi }(j - 1,m) = \frac{1}{{\sqrt M }}\sum\limits_x {T(x)} {\varphi _{j - 1,m}}(x)
\end{equation}
\begin{equation}
\label{series}
{W_\varphi }(j - 1,m) = \frac{1}{{\sqrt M }}\sum\limits_x {T(x)} {\psi}_{j - 1,m}(x)
\end{equation}
 where $W_\varphi({j-1},m)$ and $W_\psi({j-1},m)$ denote the  approximation and detail coefficients, respectively. The decomposed coefficients are combined in the following way for exact reconstruction. 
  \begin{equation}
  \label{wavelet}
  T(x) = \frac{1}{{\sqrt M }}\sum\limits_m {{W_\varphi }(j - 1,m)} {\varphi _{j - 1,m}}(x) + \frac{1}{{\sqrt M }}\sum\limits_m {{W_\psi }(j - 1,m)} {\psi _{j - 1,m}}(x)
  \end{equation}
To generate the approximation and detail coefficients, the following convolutions and downsampling can be performed employing scaling and wavelet function coefficients.
\begin{equation}
{W_\varphi }(j - 1,g) = c( - x)*T(x){|_{x = 2g,g\in\mathbb{Z^*}}}
\end{equation}

\begin{equation}
 \label{scaling}
{W_\psi }(j - 1,g) = d( - x)*T(x){|_{x = 2g,g\in\mathbb{Z^*}}}
\end{equation}

\begin{figure}[!h]
\centering
\includegraphics[scale =0.7]{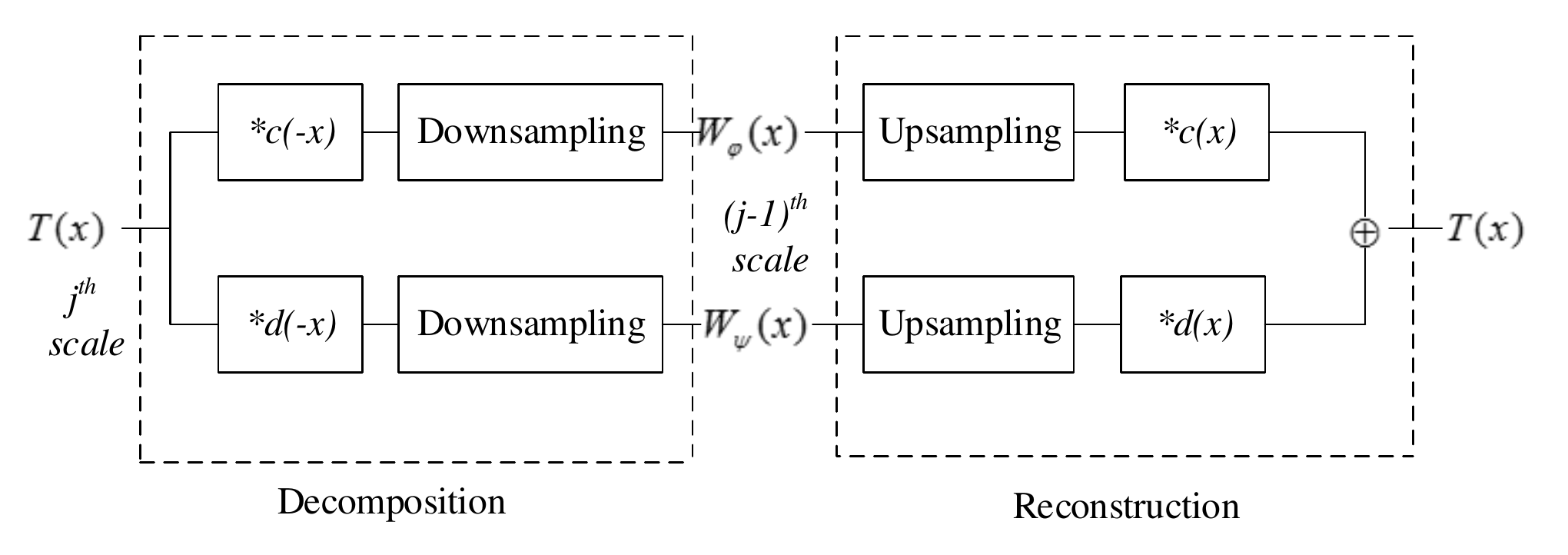}
\caption{One level decomposition and reconstruction of a 1D signal through discrete wavelet transform, with $x=2g$ evaluated only at $g \in \mathbb{Z^*}$, that is, ${W_\varphi }(j-1,g)={W_\varphi }(j-1,x)|_{g\in\mathbb{Z^*}}$ and ${W_\psi }(j-1,g)={W_\psi }(j-1,x)|_{g\in\mathbb{Z^*}}$}
\label{fig_wavelet}
\end{figure}

Just as the scaling and wavelet function coefficients are used for the analysis above, they can be used for synthesis as well to reconstruct the signal. The diagram \ref{fig_wavelet} depicts the one level decomposition (analysis) followed by reconstruction (synthesis) procedure.

The procedure applied on $T(x)$ can be further applied on the approximation coefficients at scales $j-1, j-2$ till $1$, if required, to obtain their approximation and detail coefficients. This suggests that $T(x)$ is considered the approximation co-efficient at the finest scale, that is, $T(x)={W_\varphi }(j,x)$.

As the presence of downsampling and upsampling results in loss of translation invariance, in stationary wavelet transform (SWT) \cite{nason1995stationary}, exactly the same procedure of decomposition is followed, but the downsampling and upsampling are not performed.

The above discussion  considers signals of single dimension (e.g. time series). To handle  2-dimensional signals like images, scaling and wavelet functions are required  for both row-wise and column-wise operations. This leads us to 
\begin{equation}
\begin{array}{l}
\varphi (x,y) = \varphi (x)\varphi (y)\\
{\psi ^{HL}}(x,y) = \psi (x)\varphi (y)\\
{\psi ^{LH}}(x,y) = \varphi (x)\psi (y)\\
{\psi ^{HH}}(x,y) = \psi (x)\psi (y)
\end{array}
\end{equation}
where $\varphi (x,y)$ and $\psi^i (x,y), i\in{HL,LH,HH}$ are the two dimensional scaling and wavelet functions, respectively, which are assumed to be separable functions. Note  that the wavelet functions are directionally sensitive as well. The wavelet  decomposition and reconstruction  for a 2-dimensional signal is defined in (\ref{eq1}), (\ref{eq2}), (\ref{eq3}) which are the 2-dimensional extension of 1-dimensional representations defined in (\ref{series}),(\ref{wavelet}),(\ref{scaling}) respectively.

\section{Particle Swarm Optimization}
\label{app2}

\RestyleAlgo{boxruled}
\begin{algorithm}[!h]
	\KwIn {Fitness function $F$ to be maximized\;
	    Position initialization $  \boldmath {X^i} \in \mathbb{R}{^D}|\forall {x} \in [x_{min},x_{max}]$\;
		Velocity initialization $  \boldmath {V^i} \in \mathbb{R}{^D}|\forall {v} \in [v_{min},v_{max}]$\;
		Initialize $P^i_{best}:=X^i$\;
		Initialize cognitive parameter $c_1$, social parameter $c_2$, inertia weight $w$\;
		Initialize $G_{best}:=argmax_{{P^1_{best},P^2_{best},..., P^i_{best},...,P^N_{N,best}}}F$\;}
\KwOut {$  \boldmath {G_{best}} \in \mathbb{R}{^D}$}
	
	\While{termination criteria not reached}{
		
			\For{$i \gets 1$ \textbf{to} $N$} {
			\If{$f(X^i)>F(P	^i_{best})$}{
		$P^i_{best}:=X^i$
		}\
		\If{$f(X^i)>F(G)$}{
		$G_{best}:=X^i$
		}\
		}
		\For{$i \gets 1$ \textbf{to} $N$} {
		\For{$d \gets 1$ \textbf{to} $D$}{
${v^{i,d}} = w.{v^{i,d}} + {c_1}.{rand()}.({p^{i,d}_{best}} - {x^{i,d}}) + {c_2}.{rand()}.({G^d_{best}} - {x^{i,d}})$
${x^{i,d}} = {x^{i,d}} + {v^{i,d}}$		
		}
		}
\	
		
	}
	
	\caption{{\sc Particle Swarm Optimization} }
	\label{algo:algorithm pso}
\end{algorithm}

Particle swarm optimization (PSO) \cite{eberhart1995new} acquires its inspiration from movement of the flock of birds searching for food and is very good at approximating solutions  using relatively less sensitive      and limited number of parameters compared to other such heuristic algorithms   in  a computationally inexpensive manner  \cite{eberhart1995new,eberhart1998comparison, lee2006application, omran2004particle}. 
 Particle's initial position and velocity are initialized either based on the prior knowledge of the problem or randomly. Movement of a particle in search space is evaluated based on two criteria. The first criterion is, personal experience, which is represented by the best value achieved by every individual particle and is termed as personal best ($P_{best}$).  The second criterion is, neighboring particle (including own) experience, which is represented by best value among all personal bests obtained so far in the population, and is known as global best ($G_{best}$). In each iteration, position and corresponding velocity of each particle  are updated based on its previous velocity, $P_{best}$ and $G_{best}$. 

The generic procedure of particle swam optimization (PSO) is shown in Algorithm \ref{algo:algorithm pso}.  In a population   of  N particles, the corresponding $i^{th}$ particle position with respect to a D-dimensional search is defined as  ${X^i} = \{ {x^{i1}},{x^{i2}},..........,{x^{iD}}\} $ and velocity of the particle is defined as ${V^i} = \{ {v^{i1}},{v^{i2}},..........,{v^{iD}}\} $. The positions of particles in each dimension are initialized     randomly in the range [$x_{min},x_{max}$ ] and the corresponding velocities in the range [$v_{min},v_{max}$ ]. A fitness function that optimizes desired criteria is employed. In every iteration, PSO computes the fitness of all solutions in the population, and updates $G_{best}$ and their $P_{best}$. The  updates take place only when the current $G_{best}$ and $P_{best}$ have  higher fitness values respectively than the previous. 
Then they are used along velocity values to stochastically update the population (or position) to be employed in the next iteration.   As mentioned by the authors \cite{eberhart1995new},  the implementation of PSO requires only primitive mathematical operators making it computationally inexpensive in terms of memory and speed.  But the technique may get trapped in local minima for a non-convex problem especially with limited particles and iterations.

It  is straightforward that in our use of PSO, the weights $W_{1-6}$ form 6-dimensional search space from where the $X^i$ of Algorithm \ref{algo:algorithm pso} takes values, and the fitness function is PSNR. In our algorithm, we have set the  values of  $c_1$, $c_2$ as 2 and $w$ as 1.

\end{document}